\documentclass[superscriptaddress,secnumarabic,twocolumn,
 amssymb,amsmath,nobibnotes,aps,prd,showkeys,showpacs]{revtex4}
\usepackage{graphicx,epsfig,amssymb,amsmath,amstext}
\usepackage{bm}
\usepackage{subfigure}
\usepackage{nicefrac}
\usepackage{siunitx}
\usepackage{booktabs}

\newcommand{\abs}[1]{\left| #1 \right|} 
\renewcommand{\d}[2]{\frac{d #1}{d #2}} 
\newcommand{\pd}[2]{\frac{\partial #1}{\partial #2}} 
 
\newcommand{\grad}[1]{\mathbf{\nabla} #1} 
\let\baraccent=\= 
\renewcommand{\=}[1]{\stackrel{#1}{=}} 

\begin{document}
\title{Gauss collocation methods for efficient structure preserving integration of post-Newtonian equations of motion}
\author{Jonathan Seyrich}
\affiliation{Mathematisches Institut, Universit\"{a}t T\"{u}bingen,
Auf der Morgenstelle, 72076 T\"{u}bingen, Germany}
\email{seyrich@na.uni-tuebingen.de}
\begin{abstract}
 In this work, we present the hitherto most efficient and accurate method for the numerical integration of post-Newtonian equations of motion. We first transform 
 the Poisson system as given by the post-Newtonian approximation to canonically symplectic form. Then we apply Gauss Runge-Kutta schemes 
to numerically integrate the resulting equations. This yields a convenient method for the structure preserving long-time integration of 
 post-Newtonian equations of motion. In extensive numerical experiments, this approach turns out to be faster and more accurate i) than previously proposed structure preserving 
splitting schemes and ii) than standard explicit Runge-Kutta methods.
\end{abstract}

\maketitle

\section{Introduction}\label{sec-I}
When Einstein gave birth to general relativity with the presentation of his field equations in 1915, new phenomena such as black holes and gravitational waves were soon predicted as
consequences of this theory. In the last couple of years, gravitational waves have attracted ever more attention. With the aim to finally receive signals of such waves, much experimental 
effort has been put upon mounting land-based detectors. \textit{Virgo} in France and Italy, \textit{GEO 600} in Germany and the UK, and \textit{LIGO} in the USA are only to name a 
few. They are soon to be joined by the space-based \textit{eLISA}. In order to track any signal of gravitational waves, templates are required that give a hint on which needle to 
look for in the haystack of data delivered by all the working detectors.
Such templates, in turn, can only be obtained by singling out the most promising sources of gravitational waves and calculating their motion in phase space. The main source of waves 
have been identified to be binary systems consisting of inspiraling compact objects, see, e.g., \cite{blanchet2002}. Their mass proportions can be anything between equal masses and extreme
 ratios. Binaries with very unequal masses are called Extreme Mass Ratio Inspirals (EMRIs). One common example of an EMRI is a neutron star that orbits a super massive black hole 
 (SMBH). EMRIs allow for a simple description as a free particle (the lighter one) moving in a curved spacetime given by the 
 metric corresponding to the mass of the heavier particle. Many possible shapes of the background metric have been proposed in this field, e.g., \cite{Kerrmetric, MSMmetric}.
 
 Binaries with not so extreme a mass ratio are suitably described by the post-Newtonian formalism. This approach was possible after Arnowitt, Deser and Misner discovered that 
 Einstein's theory can be formulated as a Hamiltonian System, \cite{ADM}. The idea is then to expand the elements of the metric tensor and
 the equations of motion of the matter in powers of the small parameter $\frac1{c^2}$, see, e.g., \cite{Schaefer1997}. This gives the Hamiltonian as a power series 
 in the small parameter, the first term of the series being the Hamiltonian for Newton's law of gravitation. The determination of the individual terms in this
 expansion is subject to current research in theoretical physics and contributions up to $3$PN order have been given in \cite{JarSchaef}. The post-Newtonian approach has even been extended 
 to a binary which is perturbed by a much lighter third body, e.g., \cite{BruegGal}. 
 
  A property of relativistic test-particles which is not known from classical mechanics is their spin. After the foundation for the treatment
 of this spin had been laid down in the 1950s, e.g., \cite{papapetrou}, the post-Newtonian formalism could be expanded to include the corresponding contributions. These comprise 
 spin-orbit as well as spin-spin interactions, see, e.g., \cite{DamourSchaefer1988, Damour2001}. With this extension, the Hamiltonian system becomes a so called Poisson 
 system.

 One important property of the post-Newtonian is that they are generally non-integrable. As a consequence, the motion described by them can exhibit chaotic traits.
 If the motion of a particular binary is chaotic, the gravitational waves emitted during its inspiral will be unpredictable, thus leaving the researchers 
 at the various wave detectors without any useful template. Hence, the investigation for chaos of a given binary system is an important task. 
 Consequently, many works have been published
 concerning this topic both in the geodesic and the post-Newtonian field, e.g., \cite{Lukes10, Han08, GopaKoenig1, CornLev, wuxie2007}. The analysis of chaos requires reliable indicators
 and, above all, numerical simulations over very long time spans. Numerical long-term analysis, in turn, relies on there being efficient and highly accurate
 integration schemes which behave well even during long-time simulations. To this aim one can make use of the post-Newtonian equations' special structure.
 
 Over the last few decades, the numerical analysis community came up with tools for the long-term integration of equations of motion. In the course of this, structure 
 preserving algorithms such as symplectic schemes for Hamiltonian systems (e.g., \cite{ruth, feng, hairerlubichwanner2003}) or symmetric integrators for time-reversible systems 
(e.g., \cite{McLachlan, hairersoederlind}) have been proposed. Regarding long-time behaviour and conservation properties, these schemes are superior to ordinary
 numerical integrators such as explicit Runge-Kutta schemes in many applications of classical mechanics and astronomy. Whereas for standard integration schemes the overall error 
 is normally proportional to the square of the length of the integration interval $t_i$, it only increases linearly with $t_i$ for structure preserving integrators.
 And whereas there is a drift in constants of motions for standard methods, these constants are conserved up to a small error over extremely long times for symplectic algorithms.
 These algorithms have been successfully applied even in quantum mechanics. A comprehensive
 presentation of such methods is given by \cite{hairerlubichwanner}. 

 In the last years, attempts have been made to construct structure preserving integrators for relativistic systems of compact binaries. Most recently, such a scheme has been
 proposed for the geodesic approximation, see \cite{seyrichlukes}. In the realm of post-Newtonian equations, two different approaches have been considered so far. First, a non-canonically
 symplectic integrator has been constructed which preserves the system's Poisson structure, see \cite{lubich}. Then, a transformation to canonical form has been proposed, see \cite{wuxie},
 after which symplectic methods have been applied, \cite{zhongwu}. All the previous  approaches have in common that they are based on a splitting of the Hamiltonian into a
 Newtonian part and other relativistic contributions. In this work, we will first argue and then demonstrate via numerous experiments that a more efficient and accurate method for the 
solution of post-Newtonian equations consists of a transformation to symplectic form followed by the application of Gauss Runge-Kutta schemes. This will drastically reduce the numerical
 effort when simulating post-Newtonian systems.

Our work is organized as follows: We first explain our notation in Section~\ref{sec-not}. Afterwards we 
will discuss the post-Newtonian equations of motion and their numerical properties in Section~\ref{sec-II}. In Section~\ref{sec-III} we briefly
summarize the main aspects of the Poisson integrator of \cite{lubich}. Section~\ref{sec-IV} deals with the transformation to canonical form. The subsequent Section~\ref{sec-V} presents
common splitting methods. Then, in Section~\ref{sec-VI} we present Gauss Runge-Kutta schemes and argue why they are a good choice in post-Newtonian simulations. Finally, 
we subject the 
individual methods to extensive tests and compare them to standard explicit schemes in Section~\ref{sec-VII} before we summarize our main results in Section~\ref{sec-VIII}.            
 
 \section{Notation}\label{sec-not}
In this work we use canonically conjugate position and momentum variables and restrict ourselves to the center-of-mass frame so that we have the relevant variables 
(with $a=1,2$ denoting the individual compact objects) $\mathbf x=\mathbf x_2-\mathbf x_1$, $\mathbf p=\mathbf p_1=-\mathbf p_2$, $\mathbf S_1$ and $\mathbf S_2$.
For the sake of shorter notation we will combine the relevant variables of the phase space into one variable $\mathbf y=(\mathbf p,\mathbf x,\mathbf S_1,\mathbf S_2)^T$.
With this abbreviation we can write the equations of motion as
 \begin{align}\label{not-ode}
   \d {\mathbf y}t=f(\mathbf y),\\
   \mathbf y(0)=\mathbf y_0,
 \end{align}
 with an appropriate function $f$.
 An exact solution of a differential equation of the form~\eqref{not-ode} which starts at a given point $\mathbf y_0$ and propagates the system over a time $t$ will be denoted by
 \begin{align}
	\mathbf y(t)=\varphi_t(\mathbf y_0),
 \end{align}
 whereas a numerical approximated flow over a time step $h$ is written as $\Phi_h(\mathbf y_0)$. 
Consequently, for a given point of the phase space $\mathbf y_n$, the next point on the numerical trajectory is calculated as
 \begin{align}
 \mathbf y_{n+1}=\Phi_h(\mathbf y_n).
 \end{align}
 Finally, $I$ is a unit matrix of appropriate dimension and $J$ is the symplecticity matrix
 \begin{align}
	J=\begin{pmatrix}0& I\\-I&0\end{pmatrix}.
 \end{align}
 
\section{Post-Newtonian equations of motion} \label{sec-II}
 We consider orbital contributions up to order $3$PN as given in \cite{JarSchaef} and the leading term of the spin-orbit (SO) and the spin-spin (SS) contribution of \cite{Damour2001}, 
respectively. With all these terms, our Hamiltonian reads
 \begin{align}
   H(\mathbf p,\mathbf x,\mathbf S_1,\mathbf S_2)=&H_{\text{orb}}(\mathbf p,\mathbf x)+H_\text{SO}(\mathbf p,\mathbf x,\mathbf S_1,\mathbf S_2) \nonumber\\
 &+H_\text{SS}(\mathbf p,\mathbf x,\mathbf S_1,\mathbf S_2).
 \end{align} As the motion does not depend on the absolute value of
 the masses but on their ratio, one can without loss of generality assume the total mass $m:=m_1+m_2$ to be equal to $1$. 
Using the reduced mass $\mu:=\frac{m_1m_2}m$, $q:=\|\mathbf x\|$, $\nu:=\frac\mu m$, and the unit vector $\mathbf n:=\frac{\mathbf x}q$ and choosing units such 
that $G=c=1$, the relevant terms of the orbital Hamiltonian $H_{\text{Orb}}$ are
 \begin{align}
    H_{\text{N}}(\mathbf p,\mathbf x)&=\frac{\mathbf p^2}{2\mu}-\frac\mu q\label{eqn-PN-keplerHam},\\
    H_{1\text{PN,orb}}(\mathbf p,\mathbf x)&= \frac 1 {8\mu^3}(3\nu-1)(\mathbf p^2)^2\nonumber\\
 &-\frac1{2\mu^2q}[(3+\nu)\mathbf p^2+\nu(\mathbf n\cdot\mathbf p)^2]+\frac\mu{2q^2},\\
    H_{2\text{PN,orb}}(\mathbf p,\mathbf x)&=\frac1{16\mu^5}(1-5\nu5\nu^2)(\mathbf p^2)^3\nonumber\\
&+\frac1{8\mu^3q}\left[(5-20\nu-3\nu^2)(\mathbf p^2)^2-\nonumber\right.\\
 &\left.2-\nu^2(\mathbf n\cdot\mathbf p)^2\mathbf p^2+3\nu^2(\mathbf n\cdot\mathbf p)^4\right]\nonumber\\
 &+\frac1{2\mu q^2}[3\nu(\mathbf n\cdot\mathbf p)^2+(5+8\nu)\mathbf p^2]\nonumber\\
 &-\frac{(1+3\nu)\mu}{4q^3},\\
 H_{3\text{PN,orb}}(\mathbf p,\mathbf x)&=\frac1{128\mu^7}(-5+35\nu-70\nu^2+35\nu^3)(\mathbf p^2)^4\nonumber\\
 &+\frac1{16\mu^5q}\left[(-7+42\nu-53\nu^2-5\nu^3)(\mathbf p^2)^3\right.\nonumber\\
 &+(2-3\nu)\nu^2(\mathbf n\cdot\mathbf p)^2(\mathbf p^2)^2\nonumber\\
&\left.+3(1-\nu)\nu^2(\mathbf n\cdot\mathbf p)^4\mathbf p^2-5\nu^3(\mathbf n\cdot\mathbf p)^6\right]\nonumber\\
&+\frac1{16\mu^3q^2}\left[\vphantom{\frac12}(-27+136\nu+109\nu^2)(\mathbf p^2)^2\right.\nonumber\\
&+(17+30\nu)\nu(\mathbf n\cdot\mathbf p)^2\mathbf p^2\nonumber\\
&\left.+\frac34(5+43\nu)\nu(\mathbf n\cdot\mathbf p)^4\right]\nonumber\\
&+\frac1{\mu q^3}\left\{\left[-\frac{25}8+\left(\frac{\pi^2}{64}-\frac{335}{48}\right)\nu-\frac{23}8\nu^2\right]\mathbf p^2\right.\nonumber\\
 &\left.+\left(-\frac{85}{16}-\frac{3\pi^2}{64}-\frac{7\nu}4\right)\nu(\mathbf n\cdot\mathbf p)^2\right\}\nonumber\\
&+\frac\mu{q^4}\left[\frac18+\left(\frac{109}{12}-\frac{21\pi^2}{32}\right)\nu\right].
 \end{align}
 The leading order spin-orbit coupling can be expressed by means of the orbital angular momentum $\mathbf L=\mathbf x\times \mathbf p$ and the effective spin
 \begin{align}
  \mathbf S_\text{eff}=\left(1+\frac{3m_2}{4m_1}\right)\mathbf S_1+\left(1+\frac{3m_1}{4m_2}\right)\mathbf S_2
 \end{align}
 as
 \begin{align}
    H_\text{SO}(\mathbf p,\mathbf x,\mathbf S_1,\mathbf S_2)=2\frac{\mathbf S_\text{eff}\cdot\mathbf L}{q^3}.
 \end{align}
 The spin-spin interaction is the sum of the three following terms:
 \begin{align}
    &H_{S_1S_2}(\mathbf p,\mathbf x,\mathbf S_1,\mathbf S_2)=\frac1{q^3}\left[3(\mathbf S_1\cdot\mathbf n)(\mathbf S_2\cdot\mathbf n)-\mathbf S_1\cdot\mathbf S_2\right],\\
   &H_{S_1S_1}(\mathbf p,\mathbf x,\mathbf S_1)=\frac{m_2}{2m_1q^3}\left[3(\mathbf S_1\cdot\mathbf n)^2-\mathbf S_1\cdot\mathbf S_1\right],\\
   &H_{S_2S_2}(\mathbf p,\mathbf x,\mathbf S_2)=\frac{m_1}{2m_2q^3}\left[3(\mathbf S_2\cdot\mathbf n)^2-\mathbf S_2\cdot\mathbf S_2\right].
 \end{align}
 Given the Hamiltonian, the dynamics of the system is described by the equations
 \begin{align}
    \d{\mathbf p}t&=-\grad_{\mathbf x}H,\label{eqn-PN-poissonsystempdot}\\
    \d{\mathbf x}t&=\grad_{\mathbf p}H,\\
    \d{\mathbf S_a}t&=\left(\grad_{\mathbf S_a}H\right)\times \mathbf S_a\label{eqn-PN-poissonsystemSdot}.
  \end{align}
 These equations define a Poisson-system for $\mathbf y=(\mathbf p,\mathbf x,\mathbf S_1,\mathbf S_2)^T$, i.e.,
 \begin{align}\label{enq-PN-properties-poisson}
    \d{\mathbf y}t=B(\mathbf y)\grad H,
 \end{align}
 with
 \begin{align}
   B(\mathbf y)&=\begin{pmatrix}0&-I&0&0\\ I&0&0&0\\0&0&B_1(\mathbf y)&0\\
 0&0&0&B_2(\mathbf y)\end{pmatrix},\\
 B_1(\mathbf y)&=\begin{pmatrix}0&-S_{1z}&S_{1y}\\S_{1z}&0&-S_{1x}\\-S_{1y}&S_{1x}&0\end{pmatrix},\\
 B_2(\mathbf y)&=\begin{pmatrix}0&-S_{2z}&S_{2y}\\S_{2z}&0&-S_{2x}\\-S_{2y}&S_{2x}&0\end{pmatrix}.
 \end{align}
 A numerical scheme which preserves this special structure can be 
expected to have a benevolent long-time behaviour similar to the symplectic case, e.g., \cite{hairerlubichwanner}.
 
 Furthermore, the post-Newtonian equations are in fact a perturbed Kepler problem, the corrections to the classical motion scaling with $\frac1{c^2}$. 
 In the units $G=c=1$ this scaling is encoded in the higher orders of $\frac1q$ or $\mathbf p^2$ in the post-Newtonian terms. As $q>1$ and $\mathbf p^2<1$ in most circumstances,
 one has a Hamiltonian of the form 
 \begin{align}\label{eqn-PN-properties-pertKep}
	H=H_\text{N}+\delta\tilde H,\qquad\delta\tilde H\ll1,
 \end{align} 
 where the `larger` part can be solved analytically. Keeping in mind the just mentioned properties of the post-Newtonian equations we now present the already 
 known integration methods.

\section{Poisson integrator for the post-Newtonian equations}\label{sec-III}
 The Poisson integrator suggested by \cite{lubich} is designed to exactly preserve the structure \eqref{enq-PN-properties-poisson}. Starting with the Hamiltonian
 \eqref{eqn-PN-properties-pertKep}, the relativistic contribution $\delta\tilde H$ is first split into an orbital part $H_{\text{PN,orb}}$ and a spin term $H_\text{SO,SS}$.
 The main idea is now to further decompose the spin-orbit and spin-spin parts as
 \begin{align}
	H_\text{SO}&=H^x_\text{SO}+H^y_\text{SO}+H^z_\text{SO},
 \end{align}
 with
 \begin{align}
	H^i_\text{SO}&=\frac2{q^3}(\mathbf S_\text{eff}\cdot\hat{\mathbf e}_i)(\mathbf L\cdot\hat{\mathbf e}_i),
 \end{align}
 and 
  \begin{align}
	H_\text{SS}=H^1_\text{SS}+H^2_\text{SS}+H^3_\text{SS}+H^4_\text{SS},
  \end{align}
  with
  \begin{align}
	&H^1_\text{SS}=-\frac{\mathbf S_1\cdot\mathbf S_2}{q^3},\\
	&H^2_\text{SS}=-\frac{\mathbf S_1\cdot\mathbf S_1}{2q^3}-\frac{\mathbf S_2\cdot\mathbf S_2}{2q^3},\\
	&H^3_\text{SS}=\frac{3(\mathbf S_1\cdot\mathbf n)(\mathbf S_2\cdot\mathbf n)}{q^3},\\
	&H^4_\text{SS}=\frac{3(\mathbf S_1\cdot\mathbf n)(\mathbf S_1\cdot\mathbf n)}{2q^3}+\frac{3(\mathbf S_2\cdot\mathbf n)(\mathbf S_2\cdot\mathbf n)}{2q^3}.
  \end{align}
 The major achievement of \cite{lubich} was to find analytical solutions $\varphi^i_\text{SO}$, $\varphi^i_\text{SS}$ for the flow of each spin related part. 
 This said, a structure preserving integrator $\Phi_\text{SO,SS}$ for the spinning terms is obtained by setting 
 \begin{align}\label{eqn-PN-poiss-phiSOSS}
	\Phi_\text{SO,SS}=\varphi^x_\text{SO}\circ\varphi^y_\text{SO}\circ\varphi^z_\text{SO}\circ\varphi^1_\text{SS}\circ\varphi^2_\text{SS}
   \circ\varphi^3_\text{SS}\circ\varphi^4_\text{SS}.
  \end{align}
 Thus, if one solves the flow $\varphi_\text{N}$ of the Newtonian part analytically and uses a symplectic scheme to calculate the orbital relativistic contributions 
 $\Phi_\text{PN,orb}$, one can finally combine the three flows $\varphi_\text{N}$, $\Phi_\text{PN,orb}$ and $\Phi_\text{SO,SS}$ to obtain a structure preserving flow.
 In Section~\ref{sec-V} we will discuss how to best arrange the individual flows.   

\section{Transformation to canonical form}\label{sec-IV}
Instead of directly preserving the Poisson structure~\eqref{enq-PN-properties-poisson} we can choose another way:  
The \textit{Darboux-Lie theorem} states that for every Poisson system $\eqref{enq-PN-properties-poisson}$ one can find a transformation
  \begin{align}\label{def-trafo-pois-sym}
	\mathbf z=\Psi(\mathbf y),
  \end{align}
  such that the system in the coordinates $\mathbf z$ is locally canonical. There are two properties of the post-Newtonian equations which enable us to find such a 
transformation in this case. Firstly, the positions and momenta are already in canonical form. Therefore, a transformation $\eqref{def-trafo-pois-sym}$ only
  has to focus on the spin coordinates. Secondly, by multiplying the equations of motions of the spins $\eqref{eqn-PN-poissonsystemSdot}$ 
  with the respective spin $\mathbf S_a$, we see that 
  \begin{align}
	&\frac12\d{\|\mathbf S_a\|}t=\d{\mathbf S_a}t\cdot\mathbf S_a=0,
  \end{align}
  i.e., the length of the individual spins is a first integral. These two observations make it surprisingly easy to achieve the transformation
   to symplectic form.
  
  From the constancy of the spin-length we see that two spin variables are redundant. The post-Newtonian system can therefore be described by $N=10$ variables. 
Because of this, \cite{wuxie} proposed the use of cylindrical coordinates for the spins. Accordingly, we set
  \begin{align}
	\mathbf S_a=m_a^2\chi_a\begin{pmatrix}\rho_a\cos(\xi_a)\\\rho_a\sin(\xi_a)\\ \xi_a\end{pmatrix},
  \end{align}
  where $\chi_a$ relates the length of an object's spin to the square of its mass. The conservation of the spin-length allows for the elimination of one of the variables 
  $(\rho_a,\phi_a,\xi_a)$. Thus, we can express $\rho_a$ in terms of $\xi_a$ as
  \begin{align}
	\rho_a=\sqrt{1-\xi_a^2},
  \end{align}
  whereby the spin and thus the Hamiltonian only depend on $\phi_a$ and $\xi_a$. 
  
  In order to deduce the equations of motion for the two independent variables, we observe 
that the following equalities hold true:
  \begin{align}
	\pd H{\phi_a}&=\pd H{S_{ax}}\pd {S_{ax}}{\phi_a}+\pd H{S_{ay}}\pd {S_{ay}}{\phi_a},\label{eqn-PN-trafoproof-hphia}\\
	\pd H{\xi_a}&=\pd H{S_{ax}}\pd {S_{ax}}{\xi_a}+\pd H{S_{ay}}\pd {S_{ay}}{\xi_a}+\pd H{S_{az}}\pd {S_{az}}{\xi_a},\label{eqn-PN-trafoproof-hxia}\\
	\pd{S_{ax}}{\phi_a}&=-\rho_a \sin(\phi_a)=-S_{ay},\label{eqn-PN-trafoproof-saxphia-say}\\
	\pd{S_{ay}}{\phi_a}&=\rho_a \cos(\phi_a)=S_{ax},\label{eqn-PN-trafoproof-sayphia-sax}\\
	S_{az}&=\chi_am_a^2\xi_a\label{eqn-PN-trafoproof-saz_xi}.
  \end{align}
  For the sake of shorter notation, we assume w.l.o.g. that $\chi_am_a^2=1$ until the end of this section.
  
  Due to relation $\eqref{eqn-PN-trafoproof-saz_xi}$, we have
  \begin{align}
	\d{\xi_a}t=\d{S_{az}}t=\pd H{S_{ax}}S_{ay}-\pd H{S_{ay}}S_{ax},
  \end{align}
  where the second equality is simply the equation of motion for the $z$-component of the spin.
  Substituting $S_{ax}$ and $S_{ay}$ with the help of equations $\eqref{eqn-PN-trafoproof-saxphia-say}$ and $\eqref{eqn-PN-trafoproof-sayphia-sax}$, and then
  applying $\eqref{eqn-PN-trafoproof-hphia}$ we get
  \begin{align}
	\d{\xi_a}t=-\pd H{S_{ax}}\pd {S_{ax}}{\phi_a}-\pd H{S_{ay}}\pd {S_{ay}}{\phi_a}=-\pd H{\phi_a}.
  \end{align}
  We now consider the time-derivatives of the $x$- and $y$- components. Taking into account the equations of motion for these components, the derivatives with regard to time are
  \begin{align}
	\pd H{S_{ay}}S_{az}-\pd H{S_{az}}S_{ay}&=\d{S_{ax}}t=\pd{S_{ax}}{\xi_a}\d{\xi_a}t+\pd{S_{ax}}{\phi_a}\d{\phi_a}t,\\
	\pd H{S_{az}}S_{ax}-\pd H{S_{ax}}S_{az}&=\d{S_{ay}}t=\pd{S_{ay}}{\xi_a}\d{\xi_a}t+\pd{S_{ay}}{\phi_a}\d{\phi_a}t.
  \end{align}
  We can multiply the first equation with $\frac {\partial S_{ay}}{\partial\xi_a}$ and the second with $\frac {\partial S_{ax}}{\partial\xi_a}$ and substract the two equations. This leads to
  \begin{align}\label{eqn-PN-trafoproof-dphidtfirst}
	&\left(\pd{S_{ax}}{\phi_a}\pd{S_{ay}}{\xi_a}-\pd{S_{ay}}{\phi_a}\pd{S_{ax}}{\xi_a}\right)\d{\phi_a}t=\nonumber\\
 &\vphantom{\left(\pd{S_{ax}}{\phi_a}\pd{S_{ay}}{\xi_a}-\right)}
\pd H{S_{ay}}\pd{S_{ay}}{\xi_a}S_{az}-\pd H{S_{az}}\pd{S_{ay}}{\xi_a}S_{ay}\nonumber\\
 &\vphantom{\left(\pd{S_{ax}}{\phi_a}\pd{S_{ay}}{\xi_a}-\right)}
	-\pd H{S_{ay}}S_{az}\pd{S_{ax}}{\xi_a}S_{ax}+\pd H{S_{ax}}\pd{S_{ax}}{\xi_a}S_{az}.
  \end{align}
  Calculating the partial derivatives of the spin components with regard to the new variables on the left hand side and some of the partial derivatives on the right hand
  side, equation~$\eqref{eqn-PN-trafoproof-dphidtfirst}$ becomes
  \begin{align}
	\xi_a\d{\phi_a}t=\pd H{S_{ay}}\pd{S_{ay}}{\xi_a}\xi_a+\pd H{S_{az}}\xi_a+\pd H{S_{ax}}\pd{S_{ax}}{\xi_a}\xi_a.
  \end{align}
  Keeping in mind that $\frac{\partial S_{az}}{\partial\xi_a}=1$ and then taking use of relation $\eqref{eqn-PN-trafoproof-hxia}$, we arrive at
  \begin{align}
	\d{\phi_a}t=\pd H{\xi_a}.
  \end{align}
  All in all, the post-Newtonian equations for the ten independent variables $\mathbf z=(\mathbf p,\xi_a,\mathbf x,\phi_a)$ read
  \begin{align}
	\d{\mathbf z}t=\d{}t\begin{pmatrix}\mathbf p\\\xi_1\\\xi_2\\\mathbf x\\\phi_1\\\phi_2\end{pmatrix}=\begin{pmatrix}0&- I\\ I&0
	\end{pmatrix}\begin{pmatrix}\grad_{\mathbf p}\\\partial_{\xi_1}
	\\\partial_{\xi_2}\\\grad_{\mathbf x}\\\partial_{\phi_1}\\\partial_{\phi_2}\end{pmatrix}H,
  \end{align}
  which is to say that the system in the new variables is symplectic. What is more, the transformation is defined globally as it is nothing other than expressing
  the spins with constant length via cylindrical coordinates. As a consequence, a structure preserving algorithm for the post-Newtonian
  equations can be obtained by carrying out the global transformation to canonical form and then applying a symplectic integrator.

\section{Schemes based on splitting}\label{sec-V}
 \subsection{On splitting methods}
  It is well known,
  e.g., \cite{liao}, that, given a Hamiltonian of the form~\eqref{eqn-PN-properties-pertKep}, an integrator which is split in this natural way has a smaller local 
 error than a comparable scheme.
 More precisely, suppose we were given some second order method. 
 We could apply it with a given step size $h$ to the whole system~\eqref{eqn-PN-properties-pertKep}, thus constructing the flow $\Phi_{\text{H},h}$.
 But we could also apply the numerical scheme only to the `small` part $\delta\tilde H$ and combine this symmetrically with the flow $\varphi_N$
 of the first term in~\eqref{eqn-PN-properties-pertKep}. This would yield the second order integrators 
 \begin{align}\label{eqn-PN-splitting-split}
	\Phi_{\text{split},h}=\varphi_{N,\frac h2}\circ\Phi_{\delta\tilde H, h}\circ\varphi_{N,\frac h2},
 \end{align} 
 and
 \begin{align}\label{eqn-PN-splitting-tildesplit}
	\tilde\Phi_{\text{split},h}=\Phi_{\delta\tilde H, \frac h2}\circ\varphi_{N,h}\circ\Phi_{\delta\tilde H, \frac h2}.
 \end{align} 
 Now, if we compared the local errors, we would get
 \begin{align}
	\|\varphi_{H,h}-\Phi_{H,h}\|=\mathcal O(h^3)
 \end{align}
 for the numerical scheme applied to the whole system, but
 \begin{align}\label{eqn-PN-integrators-error-estimate-delta}
	\|\varphi_{H,h}-\Phi_{\text{split},h}\|=\mathcal O(\delta h^3),\\
	\|\varphi_{H,h}-\tilde\Phi_{\text{split},h}\|=\mathcal O(\delta h^3),
 \end{align}
 for the splitting schemes. From this we observe that splitting can reduce a scheme's local error. 

 To see which of the two splitting methods is the better option, we first notice that for post-Newtonian equations, the relativistic parts are non-separable, 
 i.e. the Hamiltonian cannot be splitted in the form
 \begin{align}
    H(\mathbf p, \mathbf x)=T(\mathbf p)+V(\mathbf x).
 \end{align}
 Unfortunately, when a system is non-separable, symplectic schemes have to be implicit, see e.g., \cite{hairerlubichwanner}. As a consequence, 
 a splitting integrator of the form~\eqref{eqn-PN-splitting-tildesplit} has to solve a system of implicit equations twice per time step whereas
 a scheme of the form~\eqref{eqn-PN-splitting-split} leads to only one implicit system per step. Thus, the splitting $\Phi_{\text{split},h}$ can be expected to be
 more efficient than $\tilde\Phi_{\text{split},h}$. Numerical experiments by \cite{zhongwu} have confirmed this so that we will
 only consider splittings of the form~\eqref{eqn-PN-splitting-split} in the following. 
 
 \subsection{On composition methods}
 The drawback of a splitting scheme is that --no matter if we choose \eqref{eqn-PN-splitting-split} or \eqref{eqn-PN-splitting-tildesplit}-- it is of second order even if the numerical
 scheme for the $\delta\tilde H$ part is of (much) higher order. This can be overcome by clever composition: If we divide the step size $h$ into smaller intervals
 $h=\alpha_1h+\alpha_2h+\alpha_3h+...$ and set for some second order method $\Phi_{2\text{nd},h}$
 \begin{align}
	\Phi_{\text{comp},h}=\Phi_{2\text{nd},\alpha_1h}\circ\Phi_{2\text{nd},\alpha_2h}\circ\Phi_{2\text{nd},\alpha_3h}\circ...,
 \end{align}
 the thus obtained scheme $\Phi_{\text{comp},h}$ will be of higher order, provided that the $\alpha_i$ satisfy specific conditions, see, e.g., \cite{hairerlubichwanner}, chapter II.
 If the underlying second order scheme $\Phi_{2\text{nd},h}$ is symplectic, $\Phi_{\text{comp},h}$, as a composition of many symplectic operations, will be so, too.

 Let us briefly state another useful fact about the implementation of composition schemes: If we choose
 the second order basic method as $\Phi_{2\text{nd},h}=\Phi_{\text{split},h}$, we have
 \begin{align}\label{eqn-merger-procedure}
	\Phi_{\text{comp},h}&=...\circ\Phi_{2\text{nd},\alpha_ih}\circ\Phi_{2\text{nd},\alpha_{i+1}h}\circ...\nonumber\\
  &=...\circ\Phi_{\text{split},\alpha_ih}\circ\Phi_{\text{split},\alpha_{i+1}h}\circ...\nonumber\\
  &=...\circ\varphi_{N,\frac {\alpha_ih}2}\circ\Phi_{\delta\tilde H, \alpha_ih}\circ\varphi_{N,\frac{\alpha_ih}2}\nonumber\\
&\vphantom{=}\circ\varphi_{N,\frac{\alpha_{i+1}h}2}\circ\Phi_{\delta\tilde H, \alpha_{i+1}h}\circ\varphi_{N,\frac{\alpha_{i+1}h}2}\circ...\nonumber\\
  &=...\circ\varphi_{N,\frac {\alpha_ih}2}\circ\Phi_{\delta\tilde H, \alpha_ih}\circ\varphi_{N,\frac{(\alpha_i+\alpha_{i+1})h}2}\nonumber\\
&\vphantom{=}\circ\Phi_{\delta\tilde H, \alpha_{i+1}h}\circ\varphi_{N,\frac{\alpha_{i+1}h}2}\circ...~.
 \end{align} 
 In the last step we could `merge` terms thanks to the group property
	\begin{align}\label{eqn-group-property}
	 \varphi_{h}\circ\varphi_{s}=\varphi_{h+s}
	\end{align}
 which is valid for every exact flow, thus reducing the numerical effort. This would not be possible if we chose $\Phi_{2\text{nd},h}=\tilde\Phi_{\text{split},h}$ instead and, 
 consequently, we found another advantage of splitting~\eqref{eqn-PN-splitting-split} over splitting~\eqref{eqn-PN-splitting-tildesplit}.  

 One of the
 most popular composition methods is the state-of-the-art Suzuki composition, \cite{suzuki},
 \begin{align}\label{eqn-PN-integrators-suzuki-comp}
	\Phi_\text{$4$th,$h$}=\Phi_{2\text{nd},\alpha h}\circ\Phi_{2\text{nd},\alpha h}\circ\Phi_{2\text{nd},\beta h}\circ\Phi_{2\text{nd},\alpha h}\circ\Phi_{2\text{nd},\alpha h},
 \end{align}
 with
 \begin{align}
	\alpha=\frac1{4-4^{\frac13}},\\
	\beta=\frac{4^{\frac13}}{4-4^{\frac13}}.
 \end{align}
 This yields a $4$th order method which is symmetric, i.e. 
 \begin{align}
	\Phi_\text{$4$th,$h$}^{-1}=\Phi_\text{$4$th,$-h$},
 \end{align}
 whenever the underlying scheme is. After all the background information on splitting and composition methods, we are now in the position to present structure preserving integration 
 schemes for the post-Newtonian equations which have been considered so far.
 
 \subsection{Splitting schemes for post-Newtonian equations}
 We will present a Poisson integrator in accordance with \cite{lubich} as well as a symplectic splitting scheme. In both cases we will use the implicit midpoint rule, already 
proposed in \cite{feng}, which for any differential equation~\eqref{not-ode} has the form
 \begin{align}\label{def-midpoint}
    \mathbf y_{n+1}=\mathbf y_n+hf\left(\frac{\mathbf y_n+\mathbf y_{n+1}}2\right).
 \end{align} 
  It is of second order and preserves symmetry and symplecticity, see, e.g., \cite{hairerlubichwanner}.
 \begin{itemize}
  \item With the work of the previous two subsections and Section~\ref{sec-IV}, we construct a Poisson integrator as follows: We use the midpoint rule to calculate the flow 
  $\Phi_\text{PN,orb}$ corresponding to the orbital relativistic contributions. Then, we use the flow corresponding to the spin related parts as given in~\eqref{eqn-PN-poiss-phiSOSS}
 and its adjoint
	\begin{align}
		\Phi^{\ast}_{\text{SO,SS}}=\Phi^4_\text{SS}\circ\Phi^3_\text{SS}\circ\Phi^2_\text{SS}\circ\Phi^1_\text{SS}\circ\Phi^z_\text{SO}\circ
 \Phi^y_\text{SO}\circ\Phi^x_\text{SO},
	\end{align}
 and symmetrically combine them with $\Phi_\text{PN,orb}$ in the form
 \begin{align}
  \Phi^\text{Poisson}_{\delta\tilde H,h}=\Phi^{\ast}_{\text{SO,SS},\frac h2}\circ\Phi_{\text{PN,orb},h}\circ\Phi_{\text{SO,SS},\frac h2}
 \end{align}
  to obtain a numerical flow for all the relativistic parts $\delta\tilde H$ in~\eqref{eqn-PN-properties-pertKep}. This numerical flow is symmetric and of second order
 as is any flow constructed in this way, see, e.g., \cite{hairerlubichwanner}, chapter V. Therefore, we can combine it with the exact flow of the Newtonian part as 
 in~\eqref{eqn-PN-splitting-split} to obtain the second order scheme
 \begin{align}
	\Phi^\text{Poisson}_{\text{split},h}=\varphi_{N,\frac h2}\circ\Phi^\text{Poisson}_{\delta\tilde H, h}\circ\varphi_{N,\frac h2}.
 \end{align}
 This said, we can apply Suzuki's composition~\eqref{eqn-PN-integrators-suzuki-comp} which yields the $4$th order symmetric Poisson integrator
 \begin{align}\label{Poisson-integrator}
	\Phi^\text{Poisson}_\text{$4$th,$h$}=\Phi^\text{Poisson}_{\delta\tilde H,\alpha h}\circ\Phi^\text{Poisson}_{\delta\tilde H,\alpha h}
\circ\Phi^\text{Poisson}_{\delta\tilde H,\beta h}\circ\Phi^\text{Poisson}_{\delta\tilde H,\alpha h}\circ\Phi^\text{Poisson}_{\delta\tilde H,\alpha h}.
 \end{align}
 
\item In order to construct a symplectic scheme, we first apply the transformation to canonical form of Section~\ref{sec-IV}. The Hamiltonian in the new variables $\mathbf z$
  is still of the form~\eqref{eqn-PN-properties-pertKep}. As a consequence, we can proceed along the lines of the two subsections above. Therefore, we apply the implicit midpoint
 rule to the whole relativistic contribution $\delta\tilde H$. This second order method can then be combined with the analytical solution of the Kepler problem, leading to
 the symplectic second order splitting scheme
 \begin{align}
	\Phi^\text{sympl}_{\text{split},h}=\varphi_{N,\frac h2}\circ\Phi^\text{midp}_{\delta\tilde H, h}\circ\varphi_{N,\frac h2}.
 \end{align}
 Again, we take use of Suzuki’s composition and arrive at the integrator
 \begin{align}\label{sympl-integrator}
	\Phi^\text{sympl}_\text{$4$th,$h$}=\Phi^\text{sympl}_{\delta\tilde H,\alpha h}\circ\Phi^\text{sympl}_{\delta\tilde H,\alpha h}
\circ\Phi^\text{sympl}_{\delta\tilde H,\beta h}\circ\Phi^\text{sympl}_{\delta\tilde H,\alpha h}\circ\Phi^\text{sympl}_{\delta\tilde H,\alpha h},
 \end{align}
 which is symplectic and of order $4$.
 \end{itemize}
 The nice ideas behind them and their mathematical bounty notwithstanding, the just presented structure preserving algorithms based on splitting methods
 are not very efficient: Even in the case we use the group property~\eqref{eqn-group-property} to `merge` terms as illustrated in~\eqref{eqn-merger-procedure} whenever it is 
 possible, the symplectic integrator $\Phi^\text{sympl}_\text{$4$th,$h$}$ is still a composition of $11$ flows, five of which can only be computed via the solution of ten-dimensional
 implicit systems. Using the Poisson integrator~\eqref{Poisson-integrator} instead, we also have to calculate the midpoint rule five times. As it is only applied to the orbital 
 motion, the implicit systems are reduced to 
$6$ dimensions. But for this we have to pay heavily because, taking everything together, we have to calculate $67$ flows during one time step, most of which are related to
 the spin contributions and require the calculations of numerous rotations, see \cite{lubich}. All these facts, which will be confirmed in the numerical experiments section below, 
 make us look for a more efficient alternative to solve the post-Newtonian equations of motion. This is where Gauss collocation methods come into play.  
 
\section{Gauss Runge-Kutta methods}\label{sec-VI}
 Gauss-Runge-Kutta methods are in fact collocation methods. Therefore, we give some background concerning these schemes.
\subsection{On collocation polynomials}
 Given an interval $[t_0,t_0+h]$, stages $0\le c_1<...<c_s\le 1$, and an initial-value problem of the form~\eqref{not-ode},
 the polynomial $u(t)$ of degree $s$, satisfying
  \begin{align}
     &u(t_0)=\mathbf y_0\label{def-colloc1},\\
     &\dot u(t_0+c_ih)=f(t_0+c_ih,u(t_0+c_ih)),\qquad i=1,...,s,\label{def-colloc}
  \end{align}
 is called a \textit{collocation polynomial}.
 In order to solve an initial-value problem by \textit{collocation}, one has to find the polynomial $u(t)$ which satisfies the 
 collocation conditions $\eqref{def-colloc1}$, $\eqref{def-colloc}$. This gives an approximate solution of the initial value problem after a time step $h$ by setting
 \begin{align}
    \mathbf y(t_0+h)_{\text{col}}&:=u(t_0+h).
  \end{align}
  A detailed introduction to collocation methods can be found in \cite{hairernorsettwanner}.
  
 It can now readily be shown, e.g., \cite{hairernorsettwanner}, that a collocation method is equivalent to an implicit $s$-stage Runge-Kutta scheme
 \begin{align}
  \mathbf y_{n+1} &= \mathbf y_n+h\sum_{i=1}^sb_if(\mathbf Y_i)\label{def-rk-y_n+1},\\
  \mathbf Y_i &=\mathbf y_n+ h\sum_{j=1}^sa_{ij}f(\mathbf Y_j)\label{def-rk-Y_i},
 \end{align}
 with coefficients chosen as
 \begin{align}
    a_{ij}=\int_0^{c_i}l_j(t)dt,\\
    b_j=\int_0^1l_i(t)dt.
  \end{align}
 Here, $l_i(t)$ denote the \textit{Lagrange-polynomials} of degree $s$,
 \begin{align}
  l_i(t)=\prod_{i\neq j}\frac{t-c_j}{c_i-c_j}.
 \end{align}
  Depending on which set of stages $0\le c_1<...<c_s\le 1$ is chosen, different collocation methods can be constructed. By setting 
 \begin{align}
    c_i=\frac 1 2(1+\tilde c_i),
 \end{align}
  with $\tilde c_i$ being the roots of the \textit{Legendre-polynomial} of degree $s$, one obtains a \textit{Gauss collocation method}. The order of this methods is 
  $\mathcal O(h^{2s})$, 
 cf. \cite{hairernorsettwanner}, which is optimal in the sense that there are no other $s$-stage one-step methods that achieve a similar high order without further numerical ruse.
 In addition, Gauss collocation methods are symplectic and time-reversible, as is proven in \citep{hairerlubichwanner}. Due to all these properties, Gauss-Runge-Kutta
 methods are quite natural candidates for the solution of non-separable Hamiltonian systems. 
 \subsection{Gauss collocation for post-Newtonian equations}
  In order to employ Gauss Runge-Kutta methods in post-Newtonian simulations, we just have to conduct the transformation~\eqref{def-trafo-pois-sym} of Section~\ref{sec-IV}
  and then apply a Gauss collocation scheme to the whole system in the new coordinates $\mathbf z$. Doing so, we will have to solve the system of implicit equations~\eqref{def-rk-Y_i}
  for the inner stage values $\mathbf Y_i$ during each time step. This system has $s\cdot10$ dimensions. Contrary to the splitting schemes, 
 we have to solve the system only once when calculating the step 
$\mathbf z_n \rightarrow \mathbf z_{n+1}$. Besides, we can drastically reduce the effort for the solution of the implicit system if we take account of the following.
\subsection{Starting approximations}\label{subsec-start-appr}
  An implicit system has to be solved iteratively. Of course, the number of iterations necessary to obtain the solution depends on the distance between the starting guesses 
  $\mathbf Y_i^0$ and the final values $\mathbf Y_i$. All the better then, if there were a fast method to obtain guesses that are very close to the final values. This is possible
  for the Gauss collocations' implicit systems: Given the inner stage values of the last step $\mathbf z_{n-1}\rightarrow \mathbf z_n$, $\mathbf Y^\text{last step}_i$, 
  we set
  \begin{align}
    \mathbf Y^0_i=\mathbf y_{n-1}+h\sum_{j=1}^s\beta_{ij}f(\mathbf Y^\text{last step}_j).
  \end{align}
 Note that this requires no additional function evaluation as $f(\mathbf Y^\text{last step}_j)$ has had to be calculated in the previous step anyway. 
 If the coefficients $\beta_{ij}$ satisfy
 \begin{align}
    \sum_{j=1}^s\beta_{ij}c_j^{k-1}=\frac{(1+c_i)^k}{k},\qquad k=1,...,s,
 \end{align}
 one has, e.g., \cite{hairerlubichwanner}, chapter VIII, 
 \begin{align}\label{eqn-start-appr-error}
 \begin{Vmatrix}\mathbf Y_i-\mathbf Y^0_i\end{Vmatrix}=\mathcal O(h^s),\qquad i=1,...,s.
 \end{align}

 The above splitting schemes, in contrast, miss any similarly good starting approximations. Referring the interested reader to \cite{homepage} where we have listed
 the coefficients $c_i$, $b_i$, $a_{ij}$ and $\beta_{ij}$ for $s=2,3,4,6$, we now move on to the numerical tests. 

\section{Numerical experiments}\label{sec-VII}

All simulations for this work were run on a Core 2 Duo E6600 machine with $2.4$GHz and 
	$4$GB RAM. The codes for the simulations have been written in \textit{c++}.
	
In this section we test and compare the following algorithms:
 \begin{itemize}
    \item Transformation to canonical form combined with Gauss Runge-Kutta methods for $s=2,3,4$. The corresponding schemes are denoted by \textit{Gauss2}, 
     \textit{Gauss3}, and \textit{Gauss4}, respectively.
    \item The symplectic splitting scheme~\eqref{sympl-integrator} which will be referred to as \textit{Symp}.
    \item The Poisson integrator~\eqref{Poisson-integrator}, abbreviated by \textit{Poiss}.
    \item The classical $4$th order explicit Runge-Kutta scheme given by the tableau
	\begin{align}\label{RK4-tableau}
		\begin{array}{c|cccc}
		0&0&0&0&0\\
		\nicefrac12&\nicefrac12&0&0&0\\
		\nicefrac12&0&\nicefrac12&0&0\\
		1&0&0&1&0\\
		\hline
		&\nicefrac16&\nicefrac13&\nicefrac13&\nicefrac16.\\
		\end{array}
	\end{align}
	Hereafter, this method will by denoted by \textit{RK4}.
    \item  The explicit Cash-Karp Runge-Kutta scheme
	\begin{align}
	\label{CK5-tableau}
\small{
		\begin{array}{c|cccccc}
		0&&&&&&\\
		\nicefrac15&\nicefrac15&&&&&\\
		\nicefrac3{10}&\nicefrac3{40}&\nicefrac9{40}&&&&\\
		\nicefrac35&\nicefrac3{10}&-\nicefrac{9}{10}&\nicefrac65&&&\\
		1&\nicefrac{-11}{54}&\nicefrac52&-\nicefrac{70}{27}&\nicefrac{35}{27}&&\\
		\nicefrac78&\nicefrac{1631}{55296}&\nicefrac{175}{512}&\nicefrac{575}{13824}&\nicefrac{44275}{110592}&\nicefrac{253}{4096}&\\
		\hline
		&\nicefrac{37}{378}&0&\nicefrac{250}{621}&\nicefrac{125}{594}&0&\nicefrac{512}{1771},\\
		\end{array}
}
	\end{align}
	as proposed by \cite{nr}, which is of order $5$ and will be abbreviated by \textit{CK5}.
 \end{itemize}
	As the most reasonable measure for the efficiency, we compare the CPU calculation times. 
 The algorithms' accuracy is tested with the help of the relative error in the Hamiltonian
	\begin{align}\label{DeltaH}
    \Delta H=\abs{\frac{H(\mathbf y_n)-H(\mathbf y_0)}{H(\mathbf y_0)}},
	\end{align}
 and the relative error along the trajectory
	\begin{align}\label{rel-error-traj}
		err=\sqrt{\sum_{i=1}^N\left(\frac{y^i_\text{num}-y^i_\text{ex}}{y^i_\text{ex}}\right)^2}.
	\end{align}
 Here, superscript $i$ denotes a vector's $i$th component. Unless stated otherwise, the `exact` solution $\mathbf y_\text{ex}(t)$ will be given by an $s=6$-stage Gauss Runge-Kutta 
 scheme with a step size $h=0.1$ applied to the system in canonical coordinates.

	The simulations are aborted due to poor accuracy as soon as the error in the energy exceeds the tolerance
	\begin{align}\label{deltacrit}
		\Delta H>10^{-6}.
	\end{align}
	At first glance, it seems arbitrary to subject the integrators to such an upper limit on the energy error. But we will show now that such a bound is indeed necessary.
	
\subsection{On the importance of energy conservation}
\begin{figure} [htp]
  \centering
  \includegraphics[width=0.4\textwidth]{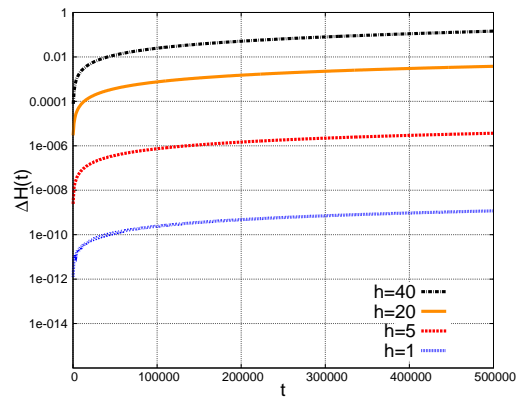}
  \caption{For the classical RK4 scheme applied with different step sizes $h$ to the purely orbital test case, 
  the error in the energy $\Delta H$ is plotted against integration time $t$.}
  \label{fig-PN-sections-RK4-energies}
 \end{figure}

\begin{figure} [htp]
  \centering
  \includegraphics[width=0.4\textwidth]{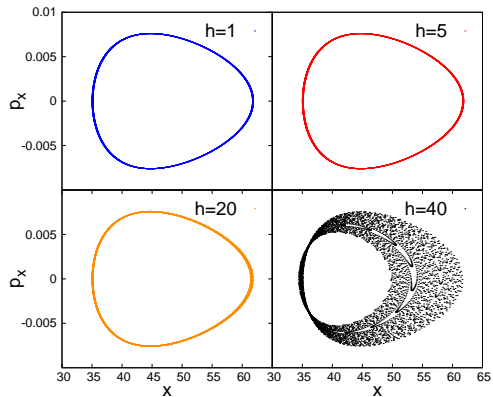}
  \caption{Poincar\'e sections at $y=0$, $p_y>0$ for the purely orbital test case obtained with RK4 and four different
  step sizes $h$.}
  \label{eqn-PN-sections-RK4-h-1-40}
 \end{figure}

Let us assume there was no upper limit on the error in the energy and we applied RK4 to the orbital test case. For different step size $h$, this would yield the energy errors as 
given in Fig.~\ref{fig-PN-sections-RK4-energies}.
 Let us now further assume we wanted to plot Poincar\'e sections for this two-dimensional problem in order to investigate it for chaotic behaviour. For different $h$, we would
 obtain the sections plotted in Fig.~\ref{eqn-PN-sections-RK4-h-1-40}. For large $h$, these resemble chaotic rather than the correct quasiperiodic motion.
 
\begin{figure} [htp]
  \centering
  \subfigure{\includegraphics[width=0.2\textwidth]{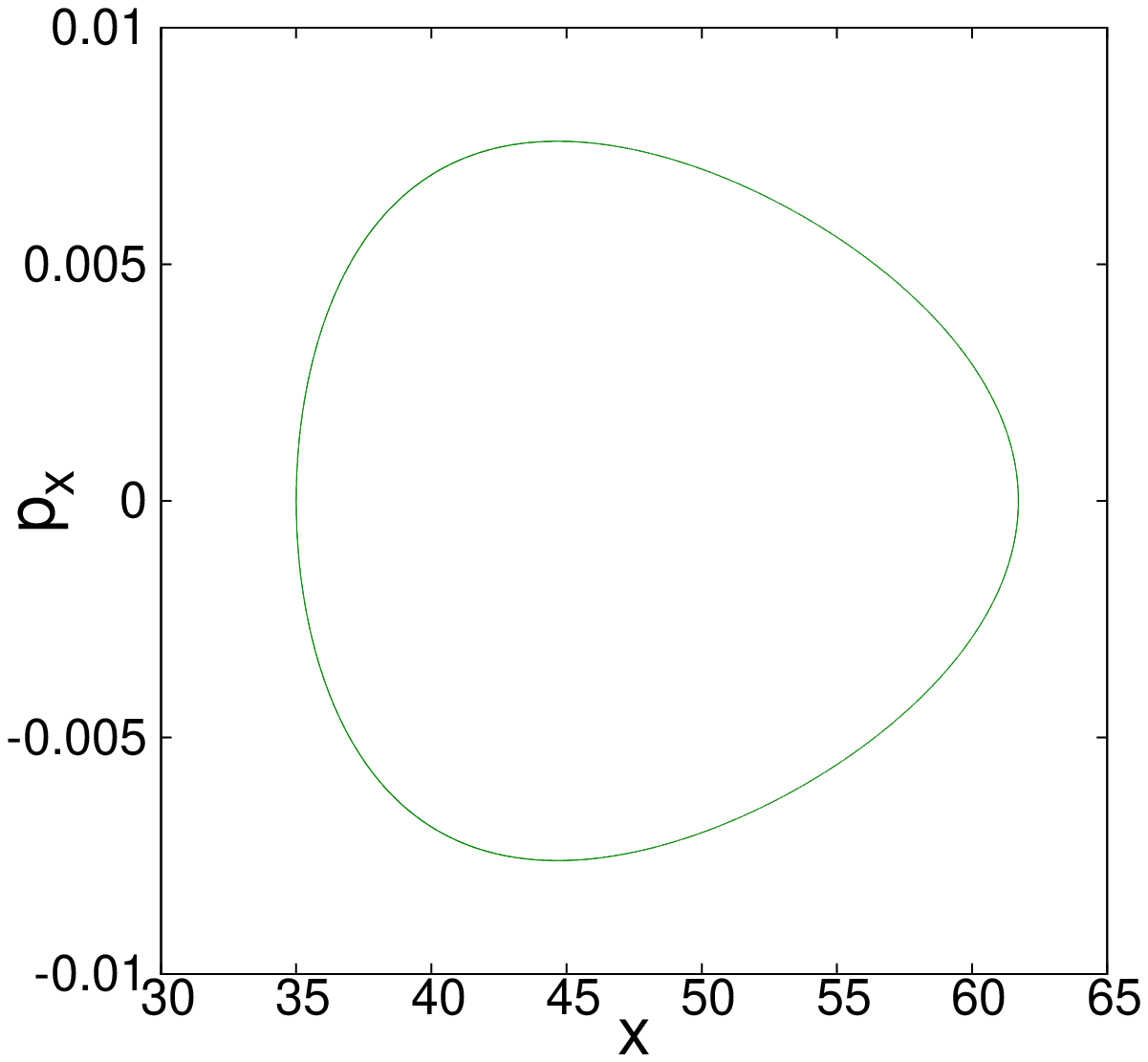}
  }
  \subfigure{\includegraphics[width=0.2\textwidth]{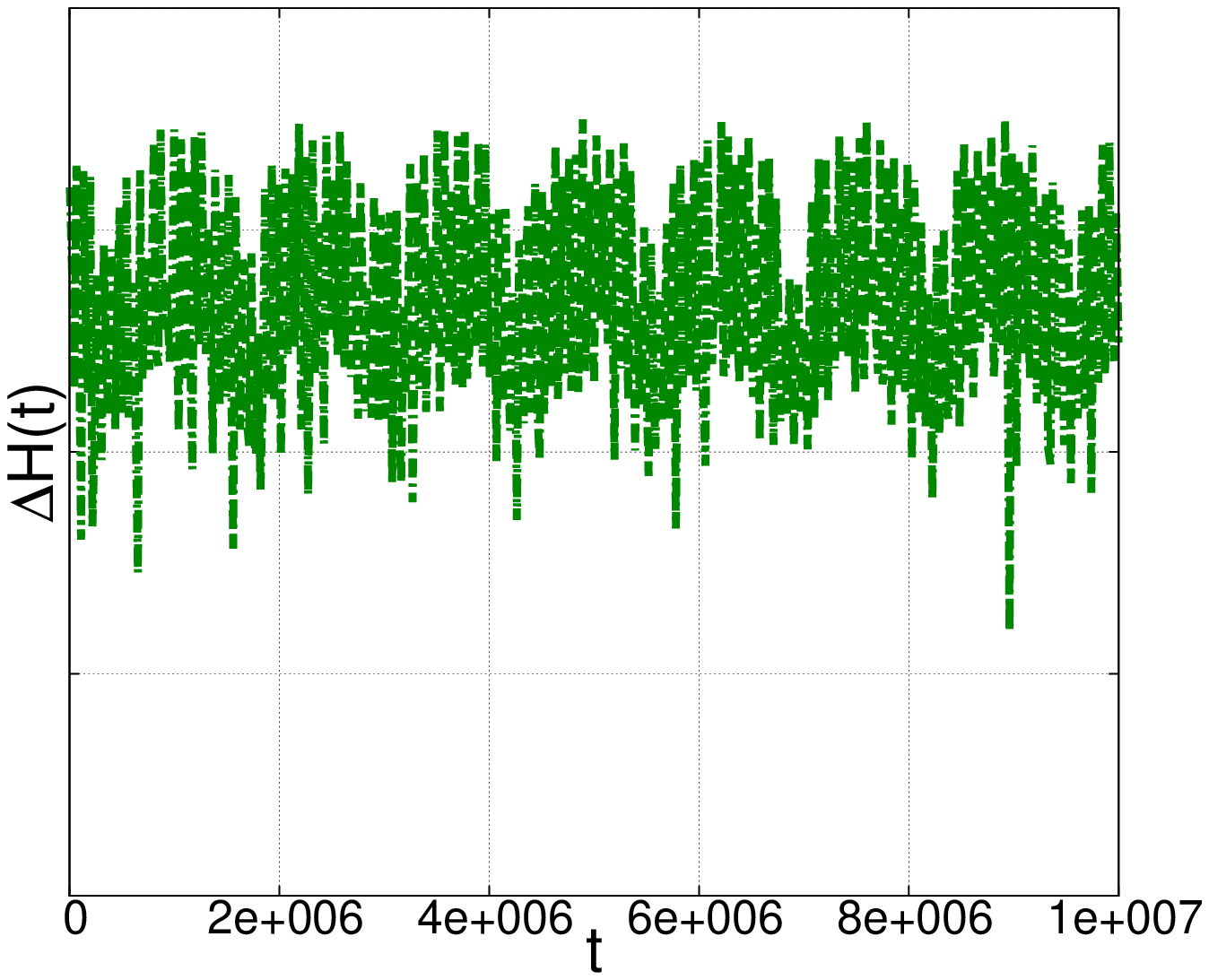}
  }
  \caption{The left panel shows Poincar\'e sections at $y=0$, $p_y>0$ for the purely orbital test case obtained with Gauss3 and $h=40$.
   In the right panel, the corresponding error in the energy $\Delta H$ is plotted against integration time $t$.}
  \label{eqn-PN-sections-gauss3-40}
 \end{figure}

Applying Gauss3 with the large step size $h=40$ instead, the energy is conserved and consequently the sections are calculated correctly, cf. Fig~\ref{eqn-PN-sections-gauss3-40}.
 We have thus illustrated that a threshold for the relative error in the energy is inevitable if we want to obtain reliable information on the chaoticity.
	Let us now present the test cases with the help of which we compare the individual methods.
	
	\subsection{The test cases}\label{subsec-PN-exp-testcases}
	We model three different kinds of motion, each of which is often encountered in binary simulations. We always fix the total mass as $m=1$. Consequently, the 
important parameter concerning the two compact object's masses is the mass ratio $\sigma=\frac{m_1}{m_2}$. The individual masses and the reduced mass are thus given as
	\begin{align}
		&m_1=\frac{\sigma}{1+\sigma},\\
		&m_2=\frac1{1+\sigma},\\
		&\mu=\frac{\sigma}{(1+\sigma)^2}.
	\end{align}
	The other relevant parameter is the factor~$\chi_a$, already introduced in section~$\ref{sec-IV}$, that links masses with spins via
	\begin{align}
		\|\mathbf S_a\|=\chi_am_a^2.
	\end{align}
	Hence, the nature of a binary's orbit depends on the parameters  $\sigma,\chi_1,\chi_2$ and the initial values
	\begin{align}
		&\mathbf z(0)=\nonumber\\
		&\left(p_x(0),p_y(0),p_z(0),\xi_1(0),\xi_2(0),x(0),y(0),z(0),\phi_1(0),\phi_2(0)\right)^T.
	\end{align}
	This said, the three kinds of motion are represented by the following respective examples:
	\begin{itemize}
		\item{}With the set of initial data
		\begin{align}
			&\mathbf z(0)=\left(0,\frac3{80},0,0,0,35,0,0,0,0\right)^T,\nonumber \\
			&\sigma=\frac13,\label{testcase-orbit}\\
			&\chi_1=\chi_2=0,\nonumber
		\end{align}
		we model a system without spin effects. The spin contributions being switched off, the post-Newtonian system is
		integrable, e.g., \cite{wuxie}, and the motion is restricted to the initial plane due to
        the conservation of the angular momentum. We present the orbit and the Poincar\'e sections 
		for $t\in[0,10^7]$ as obtained via `exact` integration in Fig.~$\ref{fig-illu-test-orbit}$. The motion is apparently quasiperiodic.
 \begin{figure} [htp]
  \centering
  \subfigure{
  \includegraphics[width=0.2\textwidth]{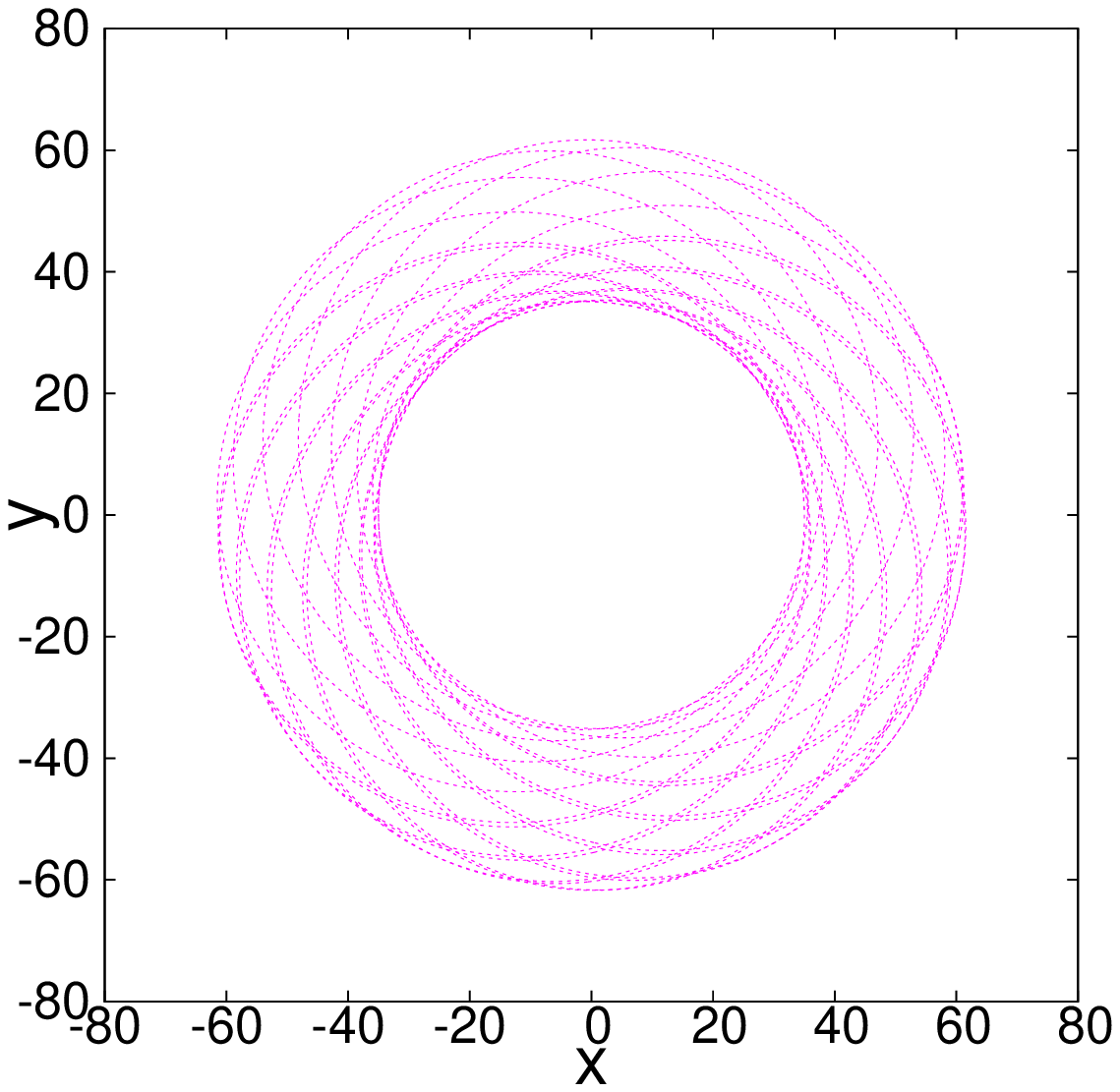}
  }
  \subfigure{
  \includegraphics[width=0.2\textwidth]{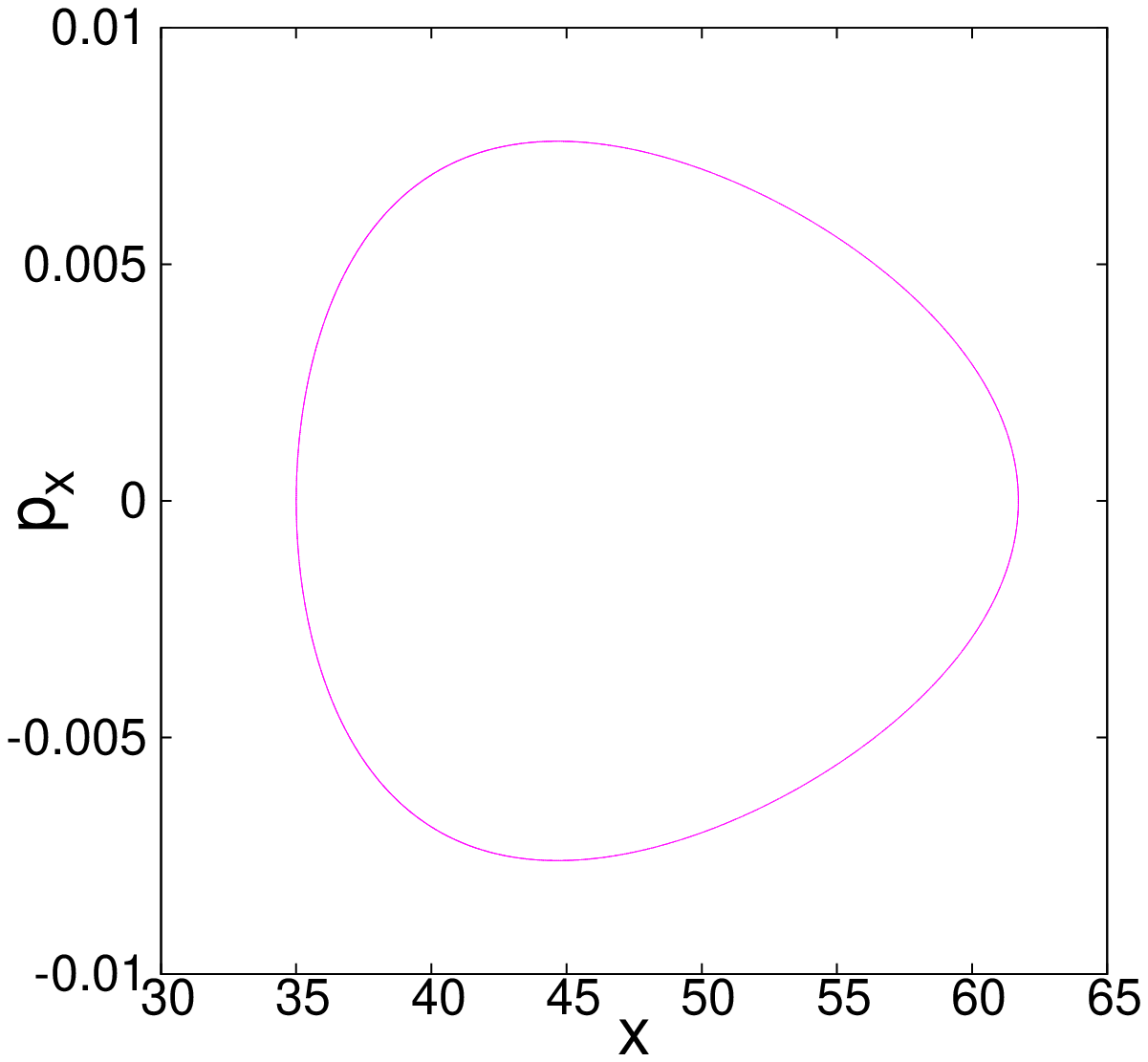}
  }
  \caption{For the test case without spin contributions and $t\in[0,10^7]$, the left panel shows an extract of the trajectory. 
   The Poincar\'e sections for $y=0$ and $p_y>0$ are given in the right panel.}
  \label{fig-illu-test-orbit}
 \end{figure} 			
		\item{}As a second test case, we choose the data set
		\begin{align}
			&\mathbf y(0)=\left(0,\frac3{80},0,0.25,-0.025,35,0,0,\frac\pi4,\frac\pi4\right)^T,\nonumber \\
			&\sigma=\frac13,\label{eqn-PN-testcase-spinreg}\\
			&\chi_1=\chi_2=\frac34.\nonumber
		\end{align}
		In Fig.~$\ref{fig-illu-test-spinreg}$, we plot a part of the orbital trajectory for $t\in[0,10^7]$. 
		Alongside this, we plot the frequency spectrum of the $x$ component for $I_1=[0,10^6]$ and $I_2=[10^7-10^6,10^7]$. We see that although the spin contributions have been 
		switched on, the motion is still regular.	
 \begin{figure} [htp]
  \centering
  \subfigure{
  \includegraphics[width=0.2\textwidth]{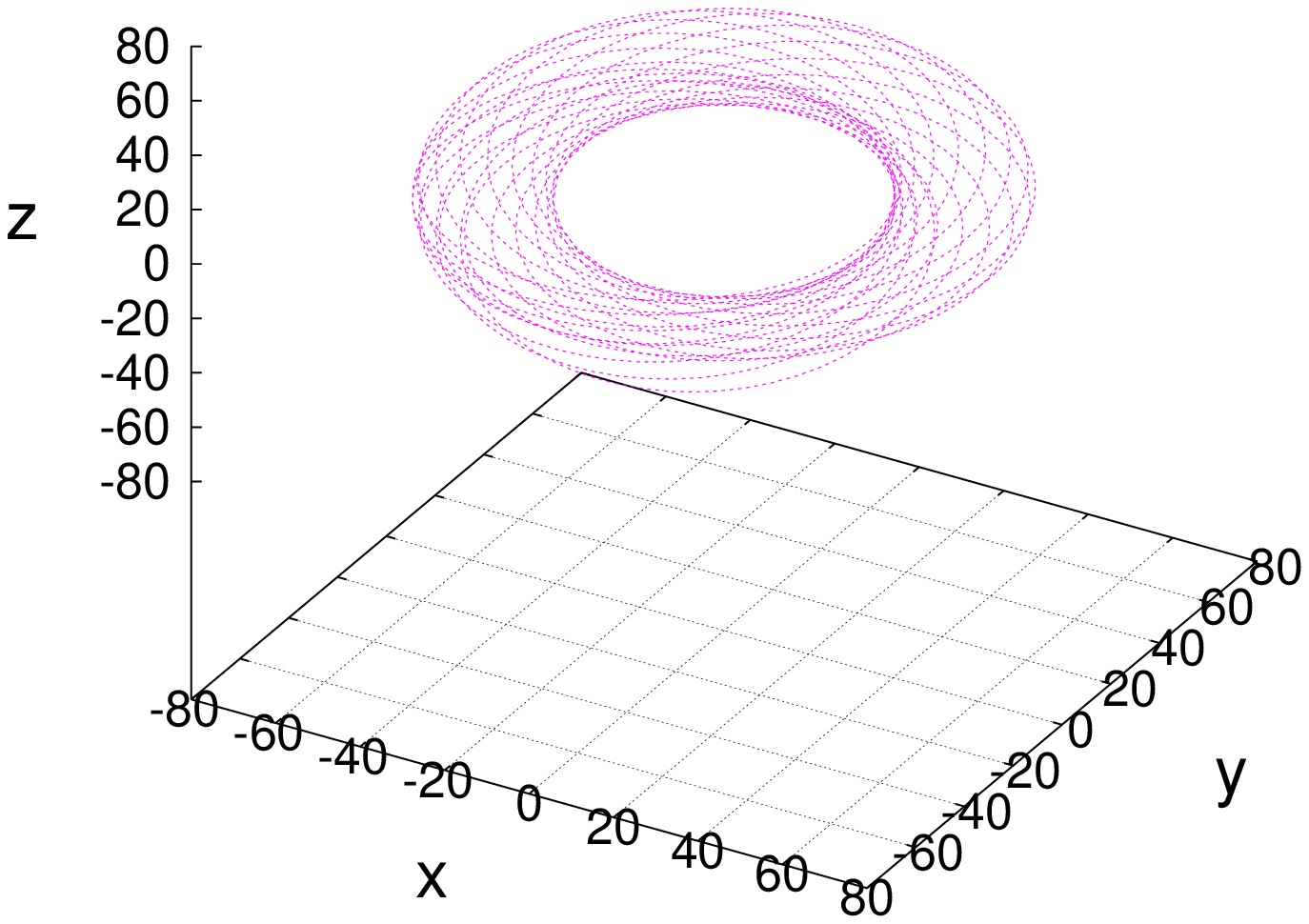}
  }
  \subfigure{
  \includegraphics[width=0.2\textwidth]{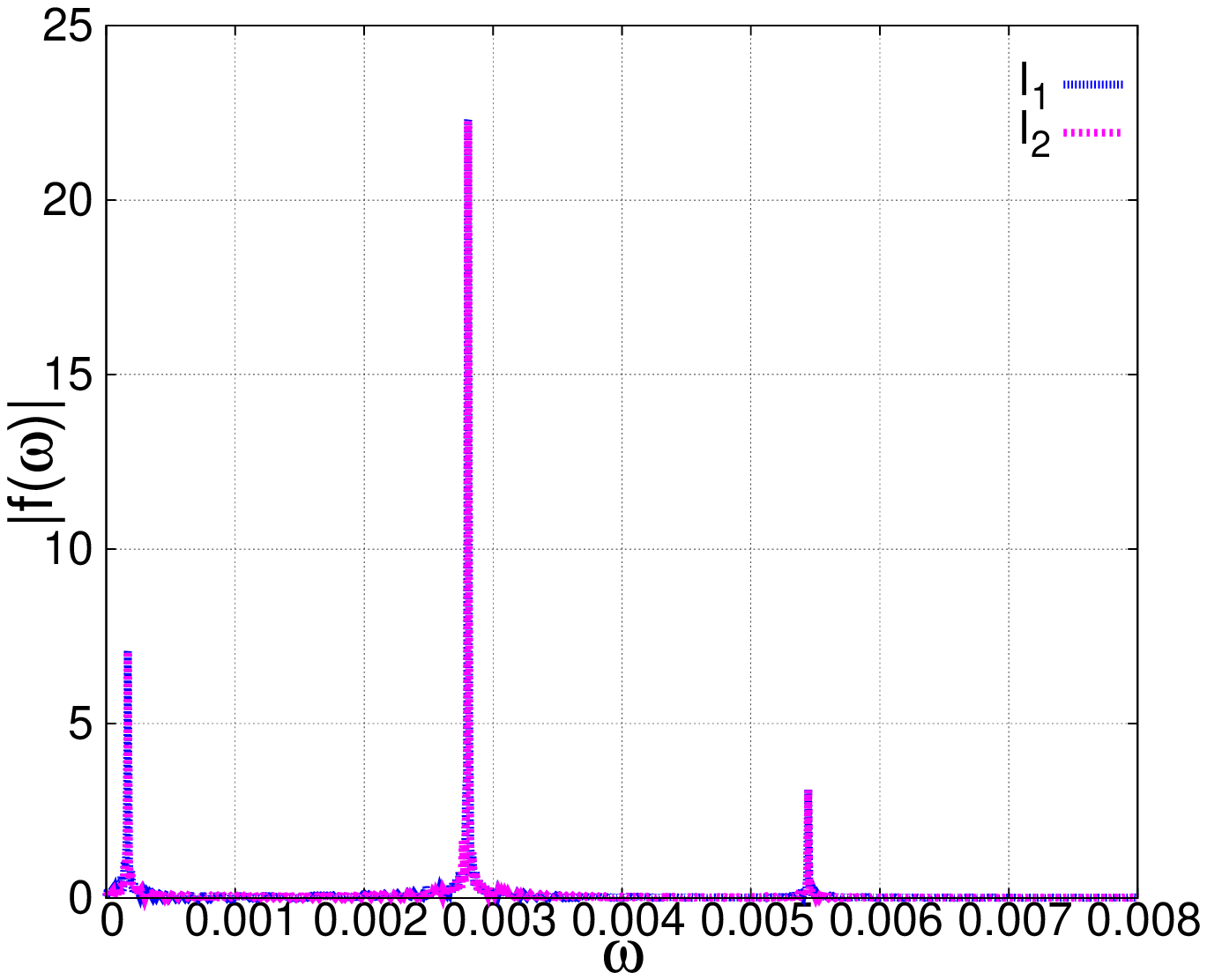}
  }
  \caption{For the test case~$\eqref{eqn-PN-testcase-spinreg}$, the left panels shows the trajectory for $t\in[0,50000]$. The frequency spectra
  $\abs{f^x(\omega)}$ for the time intervals $I_1=[0,10^6]$ and $I_2=[10^7-10^6,10^7]$ are depicted in the right panel.}
  \label{fig-illu-test-spinreg}
 \end{figure} 		
		\item{}We also consider a chaotic orbit. More precisely, we set
		\begin{align}
			&\mathbf y(0)=\left(1,0,\frac3{40},0,0.25,-0.025,6,0,0,\frac\pi4,\frac\pi4\right)^T,\nonumber \\
			&\sigma=1,\label{eqn-PN-testcase-spinchaos}\\
			&\chi_1=\chi_2=1.\nonumber
		\end{align}	
		We illustrate the chaotic behaviour by showing a part of the orbital trajectory and the FLI in Fig.~$\ref{fig-illu-test-spinchaos}$. 
		The FLI shows characteristically chaotic traits, cf. \cite{wuhuangzhang}.
 \begin{figure} [htp]
  \centering
  \subfigure{
  \includegraphics[width=0.2\textwidth]{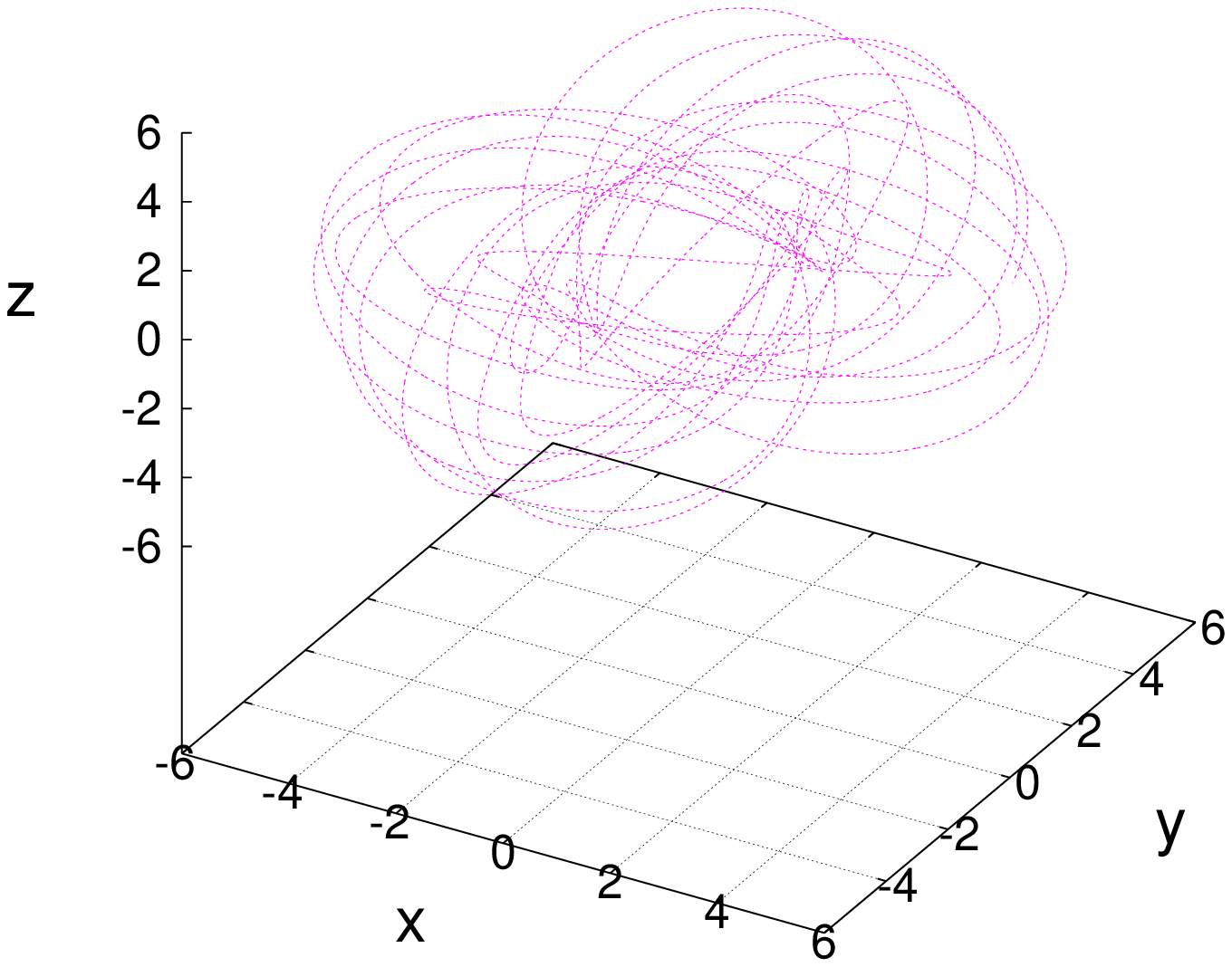}
  }
  \subfigure{
  \includegraphics[width=0.2\textwidth]{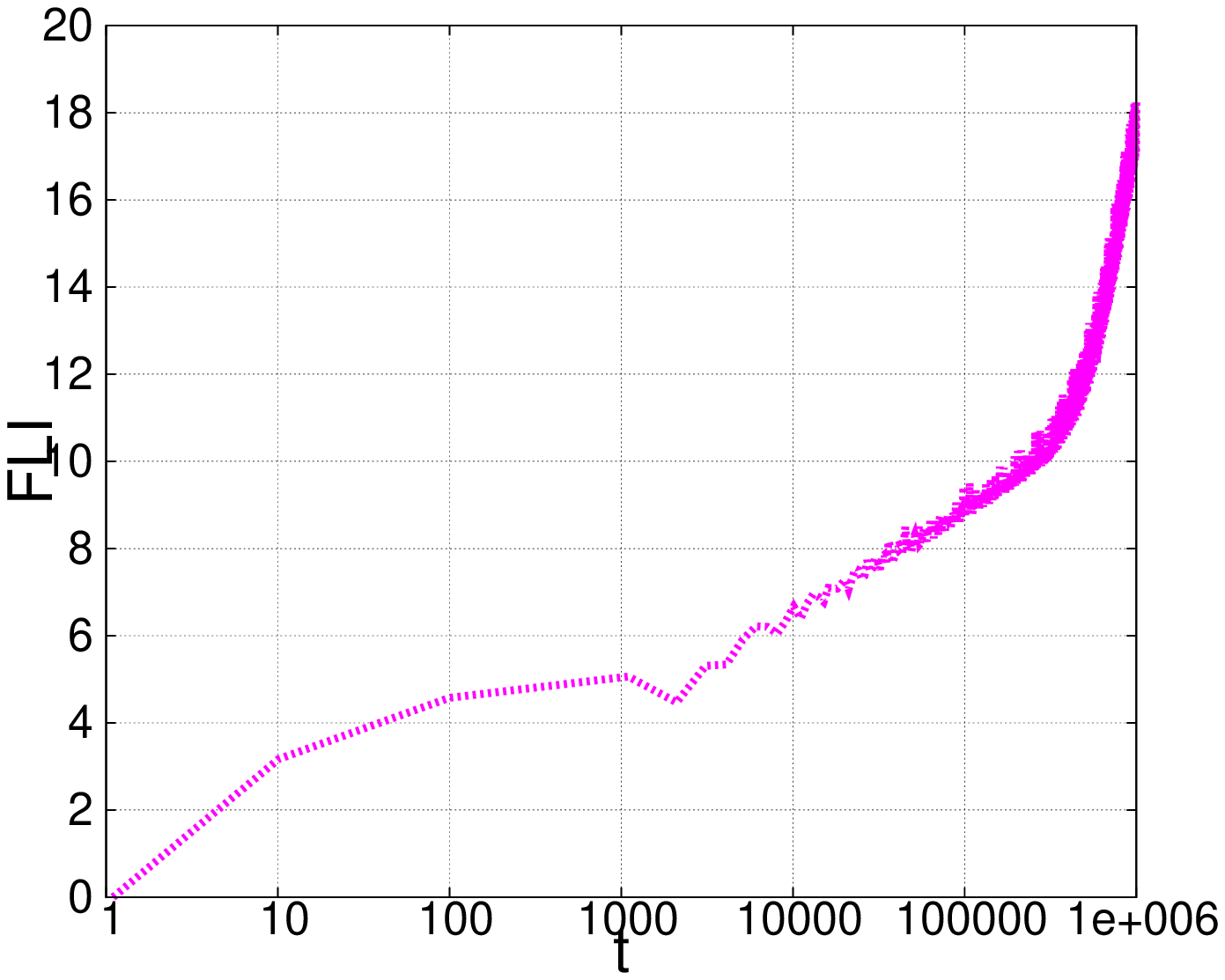}
  }
  \caption{For the chaotic test case, the left panel shows the trajectory for $t\in[0,25000]$. The linearly growing FLI is depicted in the right panel in semi-logarithmic scale.}
  \label{fig-illu-test-spinchaos}
 \end{figure} 		
	\end{itemize}
	Having thus established the test cases, we are able to start with our experiments.
	\subsection{Comparing the splitting schemes}
\begin{figure} [htp]
  \centering
  \includegraphics[width=0.4\textwidth]{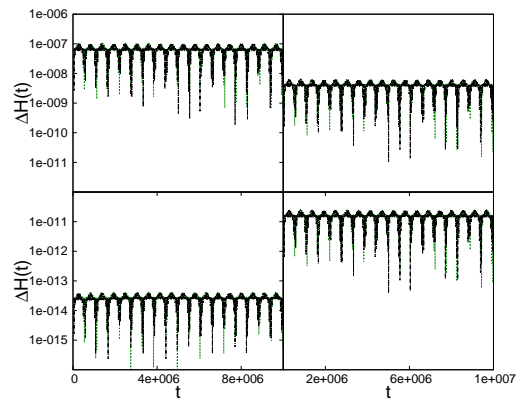}
  \caption{For initial data~$\eqref{eqn-PN-testcase-spinreg}$, $t\in[0,10^7]$ and different step sizes $h$, the relative error in the Hamiltonian $\Delta H$
  is plotted against time $t$ for the splitting integrators of section~$\ref{subsec-PN-integrators}$. No difference can be spotted between them.}
  \label{fig-compsplit-DeltaH-spinreg}
 \end{figure}

	We first compare the two splitting schemes. As they are exactly the same in the non-spinning case, we turn towards the regular spinning example~\eqref{eqn-PN-testcase-spinreg}
	and plot the respective error in the Hamiltonian for various $h$ in Fig.~\ref{fig-compsplit-DeltaH-spinreg}. We see no difference in the accuracy. But when comparing the
	corresponding calculation times in table~\ref{tab-compsplit-Tcalc-spinreg} we see that the Poisson scheme is much slower.
  \begin{table}[htp]
   \centering
  \scalebox{1.0}{
   \begin{tabular}{|c|cccc|}
    \hline
     Integrator& $h=40$&$h=20$&$h=5$&$h=1$\\
    \hline
     Symp& $9.90$&$18.72$&$67.75$&$304.95$\\
     Poiss & $17.37$&$34.23$&$133.34$&$655.11$\\
    \hline
  \end{tabular}}\vspace{4mm}
  \caption{The CPU calculation times in $[\mathrm s]$ for the two splitting integrators applied to the regular, spinning test case~$\eqref{eqn-PN-testcase-spinreg}$ 
  with different step sizes $h$. The integration
  interval was $t\in[0,10^7]$ in all simulations.}
  \label{tab-compsplit-Tcalc-spinreg}
 \end{table} 

 Testing the splitting schemes for the chaotic test case we see that Symp falls victim to criterion~\ref{deltacrit} for step sizes as small as $h=5$ but it can cope with it
 for $h<1$. Not so Poiss which even fails for $h=0.01$. Consequently, the symplectic splitting~\eqref{sympl-integrator} is superior to the Poisson 
 integrator~\eqref{Poisson-integrator}. But as we will corroborate now, it is by now means the best option for post-Newtonian systems.
 
 \subsection{Comparing integration schemes}
 Here, we compare the symplectic splitting to the (explicit and structure preserving) Runge-Kutta schemes. First, we list the calculation times for simulations with the orbital test 
 case in table~\ref{tab-CPU-gauss-exp-orbit}. As would have been expected, the explicit schemes are faster than the other methods for equal step sizes. But they have to be applied
 with small step sizes in order not to hurt the constraint on the energy error. We also see that Symp is by far the slowest algorithm.
  \begin{table}[htp]
   \centering
  \scalebox{1.0}{
   \begin{tabular}{|c|cccccc|}
    \hline
     Integrator& $h=40$&$h=20$&$h=5$&$h=1$&$h=0.5$&$h=0.1$\\
    \hline
     RK4& $\text{a}$&$\text{a}$&$\text{a}$&$13.80$&$27.58$&$137.91$\\
     CK5 & $\text{a}$&$\text{a}$&$4.70$&$23.01$&$46.07$&$230.01$\\
     Gauss2 &$\text{a}$&$3.89$&$11.44$&$43.81$&$81.73$&$344.41$\\
     Gauss3 &$3.27$&$5.32$&$15.44$&$58.48$&$105.73$&$422.69$\\
     Gauss4 &$3.96$&$6.47$&$18.74$&$67.26$&$120.59$&$443.49$\\
     Symp & $4.36$&$8.35$&$30.95$&$142.86$&&\\
    \hline
  \end{tabular}}\vspace{4mm}
  \caption{The CPU calculation times in $[\mathrm s]$ for several schemes applied to the orbital test case~$\eqref{eqn-PN-testcase-spinreg}$ 
  with different step sizes $h$. The integration
  interval was $t\in[0,10^7]$ in all simulations. `a` signifies `aborted due to condition $\eqref{deltacrit}$`.}
  \label{tab-CPU-gauss-exp-orbit}
 \end{table} 
 Doing the same observations for the regular spinning case, we get equal results, cf. table~\ref{tab-PN-CPU-gauss-exp-spinreg}.
  \begin{table}[htp]
   \centering
  \scalebox{1.0}{
   \begin{tabular}{|c|cccccc|}
    \hline
     Integrator& $h=40$&$h=20$&$h=5$&$h=1$&$h=0.5$&$h=0.1$\\
    \hline
     RK4& $\text{a}$&$\text{a}$&$\text{a}$&$43.99$&$87.93$&$439.44$\\
     CK5 & $\text{a}$&$\text{a}$&$14.31$&$71.56$&$143.02$&$716.16$\\
     Gauss2 &$\text{a}$&$10.03$&$29.37$&$111.23$&$205.68$&$852.88$\\
     Gauss3 &$8.10$&$13.19$&$38.09$&$141.19$&$255.60$&$997.65$\\
     Gauss4 &$10.00$&$16.44$&$46.63$&$160.95$&$283.26$&$1068.56$\\
     Symp& $9.90$&$18.72$&$67.75$&$304.95$&&\\
    \hline
  \end{tabular}}\vspace{4mm}
  \caption{The CPU calculation times in $[\mathrm s]$ for several schemes applied to the regular, spinning test case~$\eqref{eqn-PN-testcase-spinreg}$ with different step sizes $h$. The integration
  interval was $t\in[0,10^7]$ in all simulations. `a` signifies `aborted due to condition $\eqref{deltacrit}$`.}
  \label{tab-PN-CPU-gauss-exp-spinreg}
 \end{table} 
 
\begin{figure} [htp]
  \centering
  \includegraphics[width=0.4\textwidth]{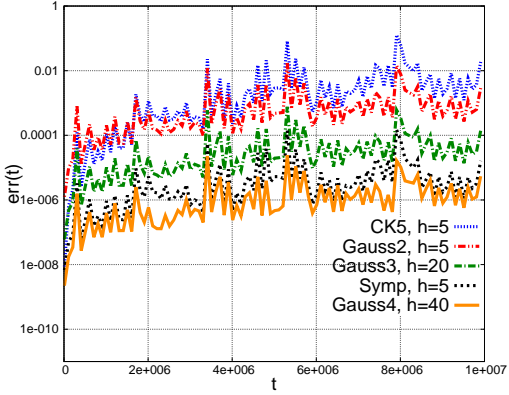}
  \caption{For initial data~$\eqref{eqn-PN-testcase-spinreg}$ and $t\in[0,10^7]$,  
  the relative error along the trajectory, cf.~\eqref{rel-error-traj}, is plotted against time $t$ for various integration schemes.}
  \label{fig-rel-error-numints-spinreg}
 \end{figure}

\begin{figure} [htp]
  \centering
  \includegraphics[width=0.4\textwidth]{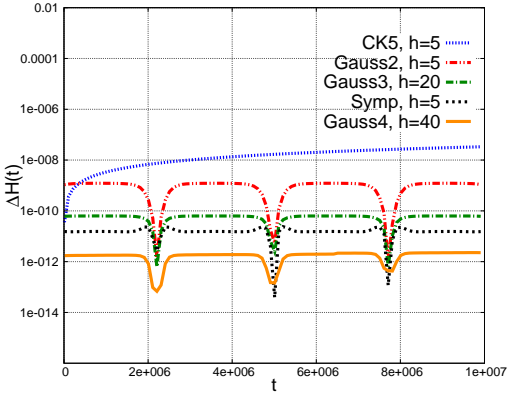}
  \caption{For initial data~$\eqref{eqn-PN-testcase-spinreg}$ and $t\in[0,10^7]$,  
  the relative error in the Hamiltonian $\Delta H$ is plotted against time $t$ for various integration schemes.}
  \label{fig-deltaH-numints-spinreg}
 \end{figure}

 As the errors of the individual integrators behave similarly for both regular orbits, we only show the case with spins included. 
 We plot the error along the trajectory~\eqref{rel-error-traj} in Fig.~\ref{fig-rel-error-numints-spinreg} and the relative error in the Hamiltonian~\eqref{DeltaH} in 
 Fig.~\ref{fig-deltaH-numints-spinreg}. We see that although Symp is more accurate than CK5 and Gauss2 with equal step sizes, it has a larger error than the much faster 
 Gauss3 and Gauss4 with equal or even much larger step sizes. 
 
\begin{figure} [htp]
  \centering
  \includegraphics[width=0.4\textwidth]{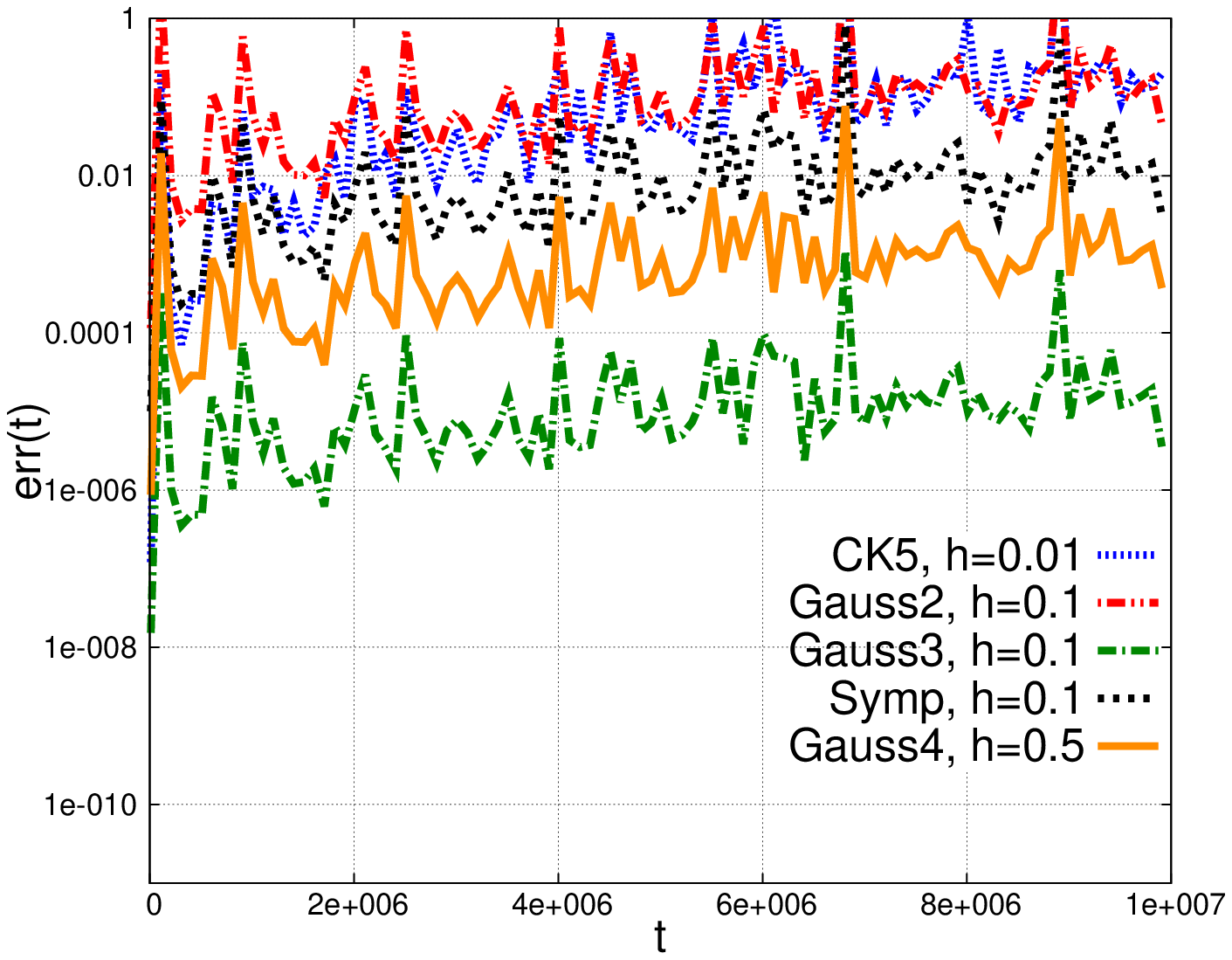}
  \caption{For initial data~$\eqref{eqn-PN-testcase-spinchaos}$,
  the relative error along the trajectory, cf.~\eqref{rel-error-traj}, is plotted against time $t$ for various integration schemes.}
  \label{fig-PN-rel-error-numints-spinchaos}
 \end{figure}

\begin{figure} [htp]
  \centering
  \includegraphics[width=0.4\textwidth]{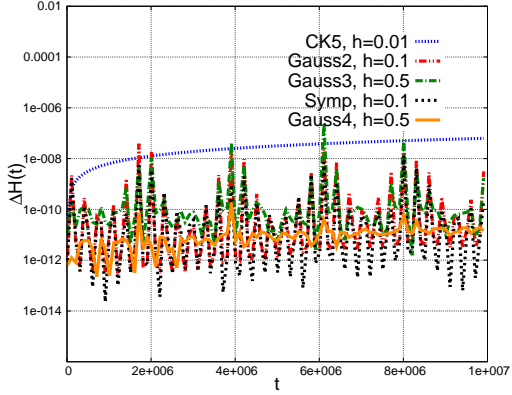}
  \caption{For initial data~$\eqref{eqn-PN-testcase-spinchaos}$ and $t\in[0,10^7]$,  
  the relative error in the Hamiltonian $\Delta H$ is plotted against time $t$ for various integration schemes.}
  \label{fig-PN-deltaH-numints-spinchaos}
 \end{figure}

 We now turn our attention towards the chaotic motion arising from the initial conditions~\eqref{eqn-PN-testcase-spinchaos}. Again, we start with listing the calculation times of
 simulations with various step sizes in table~\eqref{tab-CPU-gauss-exp-spinchaos}, right after which we plot the error along the trajectory in 
 Fig.~\ref{fig-PN-rel-error-numints-spinchaos}
 and the relative error in the energy for various simulations in Fig.~\ref{fig-PN-deltaH-numints-spinchaos}. The first point to mention here is that the explicit methods require prohibitively 
 small step sizes in order
 not to exceed the error bar~\eqref{deltacrit}. As of the structure preserving candidates, the result is qualitatively the same as in the 
 regular simulations: Symp seems to be better than Gauss2 which struggles with the chaotic case. But it obviously cannot match the performance of the fast and accurate Gauss3 and Gauss4. 
 
  \begin{table}[htp]
   \centering
  \scalebox{1.0}{
   \begin{tabular}{|c|cccccc|}
    \hline
     Integrator& $h=5$&$h=1$&$h=0.5$&$h=0.1$&$h=0.05$&$h=0.01$\\
    \hline
     RK4& $\text{a}$&$\text{a}$&$\text{a}$&$\text{a}$&$\text{a}$&$2997.76$\\
     CK5 & $\text{a}$&$\text{a}$&$\text{a}$&$\text{a}$&$\text{a}$&$4840.96$\\
     Gauss2 &$\text{a}$&$\text{a}$&$\text{a}$&$1190.14$&$$&$$\\
     Gauss3 &$\text{a}$&$\text{a}$&$449.47$&$1548.56$&$$&$$\\
     Gauss4 &$\text{a}$&$347.22$&$566.67$&$1893.45$&$$&$$\\
     Symp& $\text a$&$463.78$&$833.85$&$3445.49$&&\\
    \hline
  \end{tabular}}\vspace{4mm}
  \caption{The CPU calculation times in $[\mathrm s]$ for several schemes applied to the chaotic test case~$\eqref{eqn-PN-testcase-spinreg}$ with different step sizes $h$. The integration
  interval was $t\in[0,10^7]$ in all simulations. `a` signifies `aborted due to condition $\eqref{deltacrit}$`.}
  \label{tab-CPU-gauss-exp-spinchaos}
 \end{table} 

 One interesting point which stands out for all three initial data is that the difference in CPU times between explicit and Gauss Runge-Kutta scheme decreases for smaller step sizes.
 This is thanks to the starting approximations introduced in subsection~\ref{subsec-start-appr}. The smaller the step size, the closer the initial guess of the iterations gets to the
 correct values due to relation~\eqref{eqn-start-appr-error}. Consequently the average number of iterations per step decreases alongside $h$. To illustrate this,
 we list the iterations per step of Gauss4 in table~\ref{tab-PN-gauss-iter-per-step}. 
  \begin{table}[htp]
   \centering
  \scalebox{1.0}{
   \begin{tabular}{|c|cccccc|}
    \hline
     test case& $h=40$&$h=20$&$h=5$&$h=1$&$h=0.5$&$h=0.1$\\
    \hline
     initial values~$\eqref{testcase-orbit}$  & $9.19$&$7.44$&$5.16$&$3.46$&$2.99$&$2.13$\\
     initial values~$\eqref{eqn-PN-testcase-spinreg}$ & $9.31$&$7.59$&$5.30$&$3.54$&$3.07$&$2.24$\\
     initial values~$\eqref{eqn-PN-testcase-spinchaos}$&$$&$$&$\text{a}$&$9.21$&$7.44$&$4.85$\\
    \hline
  \end{tabular}}\vspace{4mm}
  \caption{Number of iterations per step for Gauss4, applied with different step sizes to the three test cases. `a` signifies `aborted 
  due to condition $\eqref{deltacrit}$`.}
  \label{tab-PN-gauss-iter-per-step}
 \end{table} 

 We have seen that the structure preserving algorithms have excellent conservation properties when applied to symplectic systems. What will happen if we add a radiation term to 
 the binary system?
 
 \subsection{Systems with radiation}
\begin{figure} [htp]
  \centering
  \includegraphics[width=0.2\textwidth]{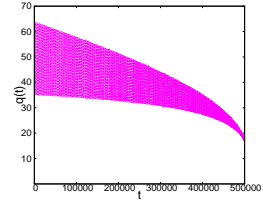}
  \caption{The radial distance $q$ as function of integration time $t$ for the regular spinning orbit with radiation effects included.}
  \label{fig-PN-illu-radiation}
 \end{figure}

 Adding a dissipative term, the system loses the structure which gave rise to the advantageous integrators in the first place. 
 But it is known from classical mechanics that,
 at least in this field, structure preserving algorithms outperform explicit schemes also when a non-conservative term is added to the Hamiltonian.
 In order to examine the corresponding behaviour for relativistic binaries, we restrict ourselves to the initial data~\eqref{eqn-PN-testcase-spinreg} and 
 modify the equation of motion of the momenta~$\eqref{eqn-PN-poissonsystempdot}$ to account for 
 radiation. We choose a model for the radiation force $F_\text{rad}$ derived by \cite{buonanno2006} which is commonly used in general relativity and set
 \begin{align}
  \d{\mathbf p}t=-\grad_{\mathbf x}H+\mathbf F_\text{rad}.
\end{align} To illustrate its effects on the trajectory, we plot the evolution of the
 radial distance $q$ for our regular, spinning test case~$\eqref{eqn-PN-testcase-spinreg}$ as given by the exact solution in Fig.~$\ref{fig-PN-illu-radiation}$. Here, 
we calculate the
`exact` solution with CK5 and the very small step size $h=0.01$. As time increases, the distance between the two particles is decreasing faster and faster. For $t>\SI{500000}{}$,
the post-Newtonian approximation will soon lose its validity. Thus, we restrict our simulations to an interval $t\in[0,\SI{500000}{}]$.

In the subsections above, CK5 and Gauss3/Gauss4 showed the best results for explicit and structure preserving schemes, respectively. We thus focus on these integrators and compare
their performance with the radiation turned on. We first list the calculation times for the three schemes applied with different step sizes each, cf. table~$\ref{tab-PN-rad-CPU}$.
With increasing time steps, the difference in CPU time becomes ever smaller as the collocation methods' average number of iterations per step decreases analogously to 
the conservative case. 

\begin{figure} [htp]
  \centering
  \includegraphics[width=0.4\textwidth]{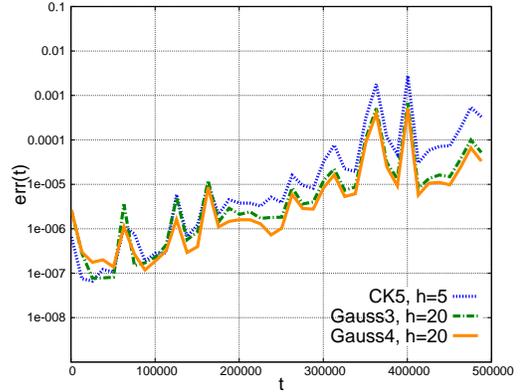}
  \caption{The relative error along the trajectory, $err$ against integration time $t$ for explicit and collocation schemes.}
  \label{fig-PN-rad-compare}
 \end{figure}

 As a measure for the accuracy we plot the relative error along the trajectory~$\eqref{rel-error-traj}$ in Fig.~$\ref{fig-PN-rad-compare}$. Taking into account the calculation
 times, the collocation methods yield the better results for less computational costs -- just as in the conservative case.
  \begin{table}[htp]
   \centering
  \scalebox{1.0}{
   \begin{tabular}{|c|cccccc|}
    \hline
     Integrator& $h=40$&$h=20$&$h=5$&$h=1$&$h=0.5$&$h=0.1$\\
    \hline
     CK5 & $0.14$&$0.28$&$1.21$&$5.75$&$11.50$&$57.51$\\
     Gauss3 &$0.92$&$1.47$&$4.04$&$16.45$&$25.69$&$99.77$\\
     Gauss4 &$1.08$&$1.72$&$4.72$&$17.28$&$32.00$&$133.36$\\
    \hline
  \end{tabular}}\vspace{4mm}
  \caption{The CPU calculation times in $[\mathrm s]$ for explicit and implicit Runge-Kutta schemes applied with different step sizes $h$ to the test case~$\eqref{eqn-PN-testcase-spinreg}$ with radiation effects included.
 The integration interval was $t\in[0,\SI{500000}{}]$ for all simulations.}
  \label{tab-PN-rad-CPU}
 \end{table}
	
\section{Summary}\label{sec-VIII} 
We have seen that structure preserving algorithms are necessary for the long-time integration of post-Newtonian equations of motion as they
guarantee the conservation of the energy which is inevitable in investigations for chaos.
Thus, in this work we analysed several algorithms -- a Poisson integrator based on the Poisson structure, a
symplectic splitting scheme and Gauss Runge-Kutta methods. We observed large discrepancies in the performance of the individual 
structure preserving methods. Some even fared worse than explicit methods. More specifically, the Poisson integrator turned out to be extremely slow when applied
to our test cases. The symplectic scheme based on state-of-the-art splitting and composition techniques could compete with a Gauss Runge-Kutta scheme with two inner stages but 
was completely
out-beaten by Gauss collocation schemes with three or more inner stages. These Gauss methods turned out to be by far the most efficient and most accurate option. Even for dissipative
systems, they delivered more accurate results for equal computational cost than high order explicit Runge-Kutta schemes. Therefore, 
we strongly recommend to use a transformation of the system to symplectic form combined with a Gauss Runge-Kutta scheme for the numerical long-time analysis of post-Newtonian
systems. 

\begin{acknowledgments}
 I would like to thank G.~Lukes-Gerakopoulos and Ch.~Lubich for useful discussions and
 suggestions. This work was supported by the DFG grant SFB/Transregio 7.
\end{acknowledgments}


\begin{thebibliography}{9}
  
  \bibitem{blanchet2002}
  L. Blanchet, {\it Living Reviews in Relativity} {\bf 5}, 3 (2002)
  
 \bibitem{MSMmetric}
  V.~S. Manko, J.~D. Sanabria-G{\'o}mez and O.~V. Manko, {\it Phys. Rev. D} 
 {\bf 62}, 044048 (2000)
 
 \bibitem{Kerrmetric}
  R.~P. Kerr and A. Schild, {\it General Relativity and Gravitation} 
 {\bf 41}, 2485 (1963)
 
  \bibitem{ADM}
  R. Arnowitt, S. Deser and C.~W. Misner, {\it General Relativity and Gravitation} {\bf 40}, 1997 (1962)
  
  \bibitem{Schaefer1997}
  G. Sch\"afer, {\it Mathematics of Gravitation} {\bf 41}, 43 (1997)
  
  \bibitem{JarSchaef}
  P. Jaranowiski and G. Sch\"afer, {\it Phys. Rev. D} {\bf 63}, 029902 (2001)
  
  \bibitem{BruegGal}
  P. Galaviz and B. Br\"ugmann, {\it Phys. Rev. D} {\bf 83}, 084013 (2011)
  
  \bibitem{papapetrou}
  A. Papapetrou, {\it Proc. R. Soc. A} {\bf 209}, 248 (1951)
  
  \bibitem{DamourSchaefer1988}
  T. Damour and G. Sch\"afer, {\it Nuovo Cimento B} {\bf 101}, 127 (1988)
  
  \bibitem{Damour2001}
  T. Damour, {\it Phys. Rev. D} {\bf 64}, 124013 (2001)

 \bibitem{Lukes10}
 G. Lukes-Gerakopoulos, T.~A. Apostolatos and G. Contopoulos, {\it Phys. Rev. D}
 {\bf 81}, 124005 (2010)

 \bibitem{Han08}
  W.-B. Han, {\it Phys. Rev. D} {\bf 77}, 123007 (2008)

 \bibitem{GopaKoenig1}
  A. Gopakumar and C. K\"onigsd\"orffer, {\it Phys. Rev. D} {\bf 71}, 024039 (2005)

 \bibitem{CornLev}
  N.~J. Cornish and J. Levin, {\it Phys. Rev. D} {\bf 68}, 024004 (2003)
  
  \bibitem{wuxie2007}
  X. Wu and Y. Xie, {\it Phys. Rev. D} {\bf 76}, 124004 (2007)
  
  \bibitem{ruth}
  R.~D. Ruth, {\it IEEE Trans. Nucl. Sci.} {\bf 30}, 2669 (1983)
  
  \bibitem{feng}
  K. Feng, {\it J. Comp. Math.} {\bf 4}, 279 (1986)
  
  \bibitem{hairerlubichwanner2003}
  E. Hairer, C. Lubich and G. Wanner, {\it Acta Numerica}, 399 (2003)
  
  \bibitem{McLachlan}
  E. Hairer, C. Lubich and G. Wanner, {\it Acta Numerica}, 399 (2003)
  
  \bibitem{hairersoederlind}
  E. Hairer and G. S\"oderlind, {\it SIAM J. Sci. Comput.} {\bf 26}, 6 (2005)

 \bibitem{hairerlubichwanner}
  E. Hairer, C. Lubich and G. Wanner, {\it Geometric numerical integration.
  Structure-preserving algorithms for ordinary differential equations}
  (Springer, 2006), 2nd ed.
  
  \bibitem{seyrichlukes}
  J. Seyrich and G. Lukes~Gerakopoulos, {\it Phys. Rev. D} {\bf 86}, 124013 (2012)
  
  \bibitem{wuxie}
  X. Wu and Y. Xie, {\it Phys. Rev. D} {\bf 81}, 084045 (2010)
  
  \bibitem{zhongwu}
  S. Y. Zhong, X. Wu, S.-Q. Liu and X.-F. Deng,
  {\it Phys. Rev. D} {\bf 82}, 124040 (2010)
  
  \bibitem{lubich}
  C. Lubich,  B. Walther and B. Brugmann,
 {\it Phys. Rev. D} {\bf 81}, 104025 (2010) 
  
  \bibitem{liao}
  X.~Liao,
 {\it Celestial Mechanics and Dynamical Astronomy} {\bf 66}, 243 (1997) 
  
  \bibitem{suzuki}
  M.~Suzuki,
 {\it Phys. Lett. A} {\bf 165}, 319 (1990) 
  
  \bibitem{hairernorsettwanner}
  E. Hairer, S. P. N\o rsett and G. Wanner,
 {\it Solving Ordinary Differential Equations I} (Springer, 1993), 2nd ed. 
  
  \bibitem{homepage}
  A file gauss\_coefficients.txt can be found on http://na.uni-tuebingen.de/\~seyrich/
  
  \bibitem{nr}
  W. Press, S. Teukolsky, W. Vetterling and B. Flannery, {\it Numerical Recipes
  in C. The art of scientific computing} (Cambridge University Press, 1992),
  2nd ed.
  
  \bibitem{wuhuangzhang}
  X.~Wu, T.~Y.~Huang, H.~Zhang
 {\it Phys. Rev. D} {\bf 74}, 083001 (2006) 
  
  \bibitem{buonanno2006}
  A.~Buonanno, Y.~Chen, T.~Damour
 {\it Phys. Rev. D} {\bf 74}, 104005 (2006) 
 
 \end{thebibliography}
\end{document}